\documentclass[a4paper,11pt]{article}
\pdfoutput=1 % if your are submitting a pdflatex (i.e. if you have
\usepackage{jcappub} % for details on the use of the package, please
\usepackage[T1]{fontenc} % if needed
\usepackage{bbold}
\usepackage{gensymb}
\usepackage{booktabs}

\usepackage[separate-uncertainty=true,
 range-units = single,
 range-phrase = -
]{siunitx}
\newcommand{\EeV}{\exa\electronvolt}

\newcommand{\be}{\begin{equation}}
\newcommand{\ee}{\end{equation}}

\newcommand{\vr}{\vec{r}}

\title{Galactic Magnetic Field Bias on Inferences from UHECR Data}

\author[a,1]{B. Eichmann,\note{Corresponding author.}}
\author[b]{T. Winchen}

\affiliation[a]{Ruhr Astroparticle and Plasma Physics Center (RAPP Center), Ruhr-Universit\"at Bochum, Institut f\"ur Theoretische Physik IV, 44780 Bochum, Germany}
\affiliation[b]{Astrophysical Institute, Vrije Universiteit Brussel (VUB), Pleinlaan 2, 1050 Brussels, Belgium \\ 
Now at Max-Planck-Institut f\"ur Radioastronomie (MPIfR), Auf dem H\"ugel 69, 53121 Bonn, Germany}

\emailAdd{eiche@tp4.rub.de}
\emailAdd{tobias.winchen@rwth-aachen.de}

\abstract{A consequence of Liouville's theorem indicates that the recently
observed large scale anisotropy in the arrival direction of Ultra-High-Energy
Cosmic Rays (UHECRs) cannot be produced by the Galactic magnetic field, thus
this anisotropy already needs to be present outside our Galaxy. But in this
case, the observed energy spectrum and composition of UHECRs differs from the
one outside of the Milky Way, due to the suppression or the amplification of
the UHECR flux from certain directions by the Galactic magnetic field. In this
work, we investigate this effect for the case of a dipole and a quadrupole
anisotropy, respectively, for the widely-used JF12 magnetic field model. We
investigate boundaries on the maximal amplitude of the observed anisotropy and
the maximal charge number of UHECRs. Furthermore, the flux modification is
discussed in the light of the Auger data on the recent dipole and also the
chemical composition. We find that this modification effect yields a
modification of the observed flux of up to $\sim 10\%$ for the investigated
magnetic field model and the observed dipole, in particular for a heavy
chemical composition of UHECRs as suggested by the 'EPOS-LHC' model.}

\begin{document}
\maketitle
\flushbottom

\section{Introduction}
\label{sec:intro}
Ultra-high energy cosmic rays (UHECRs) are believed to be charged nuclei that
penetrate Earth's atmosphere with energies above about~\SI{1}{\EeV} that are
likely accelerated in powerful extragalactic objects~\cite{Anchordoqui2019}. In
principle, this reveals the possibility for cosmic ray astronomy, however,
UHECRs get deflected by the magnetic fields inside and outside our Galaxy.
Under certain constraints --- that seem satisfied in the case of UHECRs --- the
magnetic fields cannot introduce anisotropies, according to Liouville's
theorem.  An isotropic cosmic ray distribution outside the Milky Way has to be
isotropic at Earth, and the properties of the UHECRs like the energy spectrum
do not change.  However, the arrival directions of UHECR show clear evidence of
a dipole anisotropy~\cite{Aab2017e} and  thus also the observed properties of
the UHECR flux can get modified by the Galactic magnetic field compared to the
extragalactic flux used for inferences on UHECR sources~\cite{Harari2010}.
Thus, inferences from UHECR data are impacted by the assumed magnetic fields.
Unfortunately the Galactic magnetic field is not well known and the model by
Jansson and Farrar~\cite{2012ApJ...757...14J}, hereafter referred to as JF12,
which is widely used in the cosmic ray community, fails to describe all
available data~\cite{Beck_2016, 2018ApJS..234...11H, 2019MNRAS.486.4275X}.

A first view on this effect for the JF12 model~\cite{2019ICRC...36..468W} has
shown that the modification depends significantly on the dipole direction and
its amplitude, and that especially above some tens of EeV, the change by the Galactic
magnetic field is expected to be negligible compared to the observational
uncertainties. However, in this study neither the observed change of the dipole
strength and direction, nor the increasing heaviness of the UHECR composition
or the limited field of view of the experiments have been taken into account.
This work clarifies if under consideration of the data of the Pierre Auger
Observatory, the UHECR spectrum at Earth is significantly different
from the one outside our Galaxy assuming the JF12 model and dipole as well as
quadrupole anisotropies.  Further, it discusses if this effect is able to
resolve the puzzling discrepancies of the UHECR flux~\cite{TA_Auger_ICRC2017}
between the Pierre Auger Observatory (Auger) and the Telescope Array (TA)
experiments.

\section{Method}
\label{method}
To investigate the impact of the Galactic magnetic field on UHECRs simulations
of their propagation are required.  Unfortunately, forward simulation of the
propagation through the galaxy is challenging, since the probability to hit the
Earth by chance becomes extremely small --- about $1:10^{30}$.  However, as the
propagation distance of UHECRs through the Galactic environment is
comparatively small with respect to the length scale of stochastic energy
losses, the cosmic ray energy has hardly changed from the edge of our Galaxy to
Earth. Thus, the arrival direction at Earth depends only on the initial
phase-space coordinates of the particle and its rigidity $R=E/(eZ)$ given by
the ratio of the energy $E$ over the charge $eZ$ of the particle, according to
the Lorentz force.  The effect of the magnetic field on the cosmic ray arrival
direction can thus be described as lens~\cite{Harari1999} that transforms the
cosmic ray arrival directions from outside the Milky Way to Earth.

An efficient technique to create such a Galactic lens in simulations is the
so-called backtracking method, where anti-particles are propagated backwards to
obtain the trajectories of the regular particles that hit the
Earth~\cite{2014APh....54..110B}.  Here, all trajectories that end up at the
edge of the Galaxy yield possible arrival directions at Earth.
The lens can be described as set of matrices $\mathcal{L}(\vr,\,R)$ of the
density of trajectories, that transforms the
distribution $M_0(R, \vr)$ of arrival directions $\vr$ from
outside our Galaxy to Earth dependent on the particles' rigidity $R$ to the
observed distribution at Earth $M(\vr,\,R)=\mathcal{L}(\vr,\, R)\cdot M_0(\vr)$.
To obtain matrices of finite size, the directions are binned into equal area
pixels following the HEALPix scheme~\cite{2005ApJ...622..759G}. Simulation of
the particle trajectories for the lens as well as its application to the model
maps are performed using the publicly available
CRPropa3\footnote{\url{https://crpropa.desy.de}} code~\cite{1475-7516-2016-05-038}.

In this work, the Galactic lens is generated from about $256\times
49\;152$ isotropically emitted particles for
each of 200 logarithmically binned rigidities between $0.1\,\text{EV}$ and
$10^3\,\text{EV}$. These numbers enable sufficient statistics for a resolution
of the lens of about 1 degree. Here, we use predominantly the JF12 model with
three different coherence lengths $\lambda$ of the turbulent component of the
magnetic field. Since $\lambda$ is expected to be about a fifth of the maximal
length scale of the turbulent system, we use $\lambda\in \{ 10,\, 60,\, 100
\}\,\text{pc}$.

In order to construct a certain distribution of arrival directions outside our
Galaxy, we use Healpy\footnote{\url{https://healpy.readthedocs.io}} in order to
generate a map $M_0$ of $49\;152$ pixels with a certain value. In the case of
an ideal dipole anisotropy with a dipole $\vec{d}$ these values are given by
\be M_0(\vr)=1+\vec{d}\cdot\vr\,, \ee where $\vr$ denotes the spatial direction
of a given pixel and the first term represents the normalised monopole. Note
that the dipole amplitude is given by $d=|\vec{d}|$.  In the case of an ideal
quadrupole $\tilde{Q}_{ij}=3\,q_{i}\,q_{j}-\delta_{ij}$ without any dipole
anisotropy, the distribution is determined by \be M_0(\vr)=1 +
\frac{1}{2}\sum_{i,j}\,Q_{ij}\,r_{i}\,r_{j}\,, \ee where the normalized
quadrupole $Q_{ij}$ is used, that provides an average quadrupole amplitude
$Q=\sqrt{\sum_{ij}Q_{ij}^2/9}$.

Note, that the unit vector $\vec{q}$ defines the direction of one of the two
maxima of the quadrupole distribution and the scalar $Q$ determines the
amplitude. Thus, the maximal value of the distribution $M_0$ yields $1+d$ for
the dipole and $1+Q$ for the quadrupole, respectively. We only consider a
symmetric quadrupole to keep the number of studied variables manageable, i.e.\
the eigenvalues of $Q_{ij}$ are given by $\lambda_-= \lambda_0 = -\lambda_+/2$.
So, only the strength of the quadrupole amplitude $Q$ will be varied in the
following.

In the case of an ideal detector the Galactic modification of the total UHECR
flux is given by
\begin{equation}
 \Delta F=\frac{\sum M}{\sum M_0}-1\,.
\end{equation}
Furthermore, the redistribution
of cosmic rays by the Galactic magnetic field leads to a modification of the
strength and direction of the individual components of the
anisotropy. Here, the amplitude of the dipole and the quadrupole, respectively,
is proportional to its corresponding coefficient $C_1$ and $C_2$, respectively,
of the spherical harmonics according to \be d =
\sqrt{\frac{9\,C_1}{C_0}}\,\text{ and } Q = \sqrt{\frac{50\,C_2}{3\,C_0}}\,,
\ee where $C_0$ denotes the coefficient of the monopole.  Hence, we compare the
resulting coefficient $C_l(M)$ after the lensing with the coefficient
$C_l(M_0)$ before the lensing to obtain a measure of the change of the
anisotropy amplitude. So, the modification of the anisotropy amplitude is given by
\begin{equation}
    \Delta d=\sqrt{\frac{C_1(M)\,C_0(M_0)}{C_1(M_0)\,C_0(M)}}-1\,,\, \text{ and }
    \Delta Q=\sqrt{\frac{C_2(M)\,C_0(M_0)}{C_2(M_0)\,C_0(M)}}-1\,,
\end{equation}
respectively. In addition, the regular component of the Galactic magnetic field
also changes the direction of the anisotropy. To account for this effect in
case of a dipole, we use the Galactic lens to determine the angular change
\begin{equation}
    \vec\Delta_d = \arccos\left(\vec d\cdot \vec d_{\rm obs}/|\vec d_{\rm obs}|\right)
\end{equation}
of the direction of the dipole from outside the Galaxy to Earth. In the case of
the quadrupole, the correlation of the directions of the maximas outside the
Galaxy to the ones at Earth becomes indistinct --- in particular for strong
direction changes that already occur at $R\lesssim 10\,\text{EV}$. In addition,
the lens also introduce anisotropies of different order, which makes a naive
correlation ambiguous. Therefore, we only include the change the dipole
direction in this work in particular as no quadrupole has been observed yet.

The lens $\mathcal{L}$ is build from the backtracking of anti-particles
into 49\;152 equidistant directions, so that not necessarily every direction
outside the Galaxy hits the Earth. Therefore, it is not trivial to invert the
lens $\mathcal{L}$ and we choose a more simple approach, where we build a
rigidity dependent correlation scheme that relates the corresponding dipole
direction at Earth for 3072 equidistant dipole direction outside the Galaxy.
So, for an arbitrary dipole direction at Earth, we use the one that possesses
the smallest angle with respect to this direction.

In order to account for the limited field of view of the UHECR observatories,
we use the declination $\delta$ dependent directional exposure
\cite{2001APh....14..271S}
\be
\omega(\delta) = \cos(\lambda_i)\,\cos(\delta)\,\sin(\arccos(\xi))+\arccos(\xi)\,\sin(\lambda_i)\,\sin(\delta))\,,
\ee
with
\be
\xi(\delta) = \frac{\cos(\alpha_i) - \sin(\lambda_i)\, \sin(\delta)}{\cos(\lambda_i)\, \cos(\delta)}\,.
\ee
Here, $\lambda_i$ denotes the latitude of the considered experiment, hence,
$\lambda_{\rm Auger}=-35.25\degree$ for Auger and $\lambda_{\rm
TA}=39.30\degree$ for TA. Further, $\alpha_i$ is the maximal zenith angle of
the arriving cosmic ray that is taken into account by the experiment. Although,
$\alpha_i$ slightly depends on the energy, we use a constant value of
$\alpha_{\rm Auger}=80\degree$ and $\alpha_{\rm TA}=45\degree$, respectively.
Since the observatories already account for this exposure dependence in their
spectrum measurement, we only have to consider the different field of views
given by the acceptance
\begin{equation}
    a(\delta)=\begin{cases} 4\pi/\Omega_i\,\text{ for }\omega(\delta)\neq 0\,\\
    0\,\text{ for }\omega(\delta)= 0\,,
    \end{cases}
\end{equation}
where $\Omega_i$ denotes the non-vanishing solid angle of the detector's field
of view, in order to obtain the detector dependent UHECR flux. Thus, each pixel
of the given map has a certain declination dependent acceptance $a(\delta)$, so
that $M' = M \cdot a(\delta)$, and $M'_0 = M_0 \cdot a(\delta)$ yields the
resulting distribution at the observatories with and without the impact by the
Galactic magnetic field, respectively. Hence, $\Delta F_i=\sum M'/\sum M'_0-1$
provides the Galactic modification of the total UHECR flux dependent on the
observatories field of view.

\section{General bias}
\label{results}
In the following, we use the JF12 model of the Galactic magnetic field with a
turbulent coherence length of $\lambda=60\,\text{pc}$, unless otherwise stated.
Hereby, we evaluated 3072 equidistant anisotropy directions at Earth which
yields a resolution of $<5\degree$.

\subsection{Anisotropy amplitude}
\begin{figure}[htbp]
\centering
\includegraphics[width=.32\linewidth]{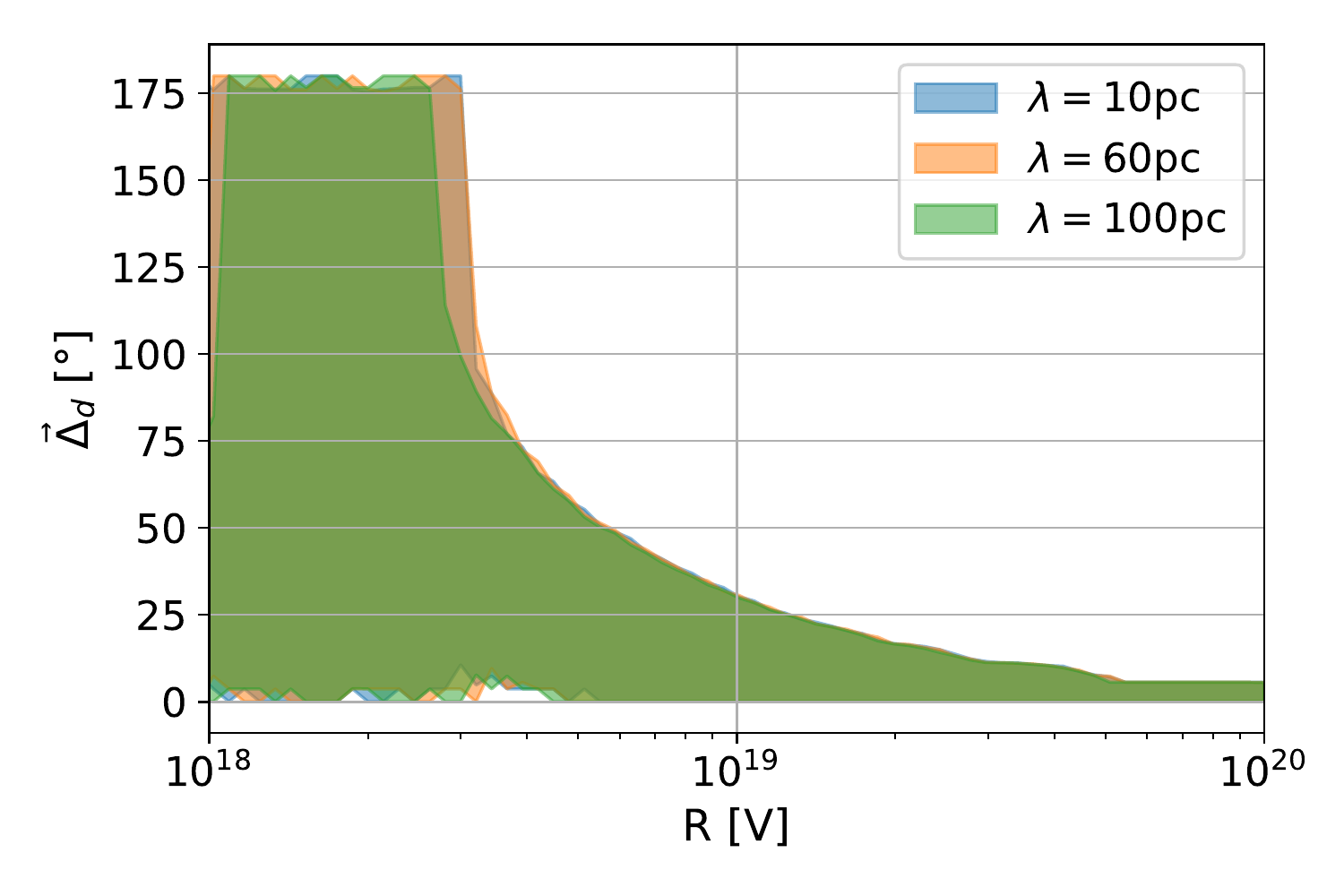}
\includegraphics[width=.32\linewidth]{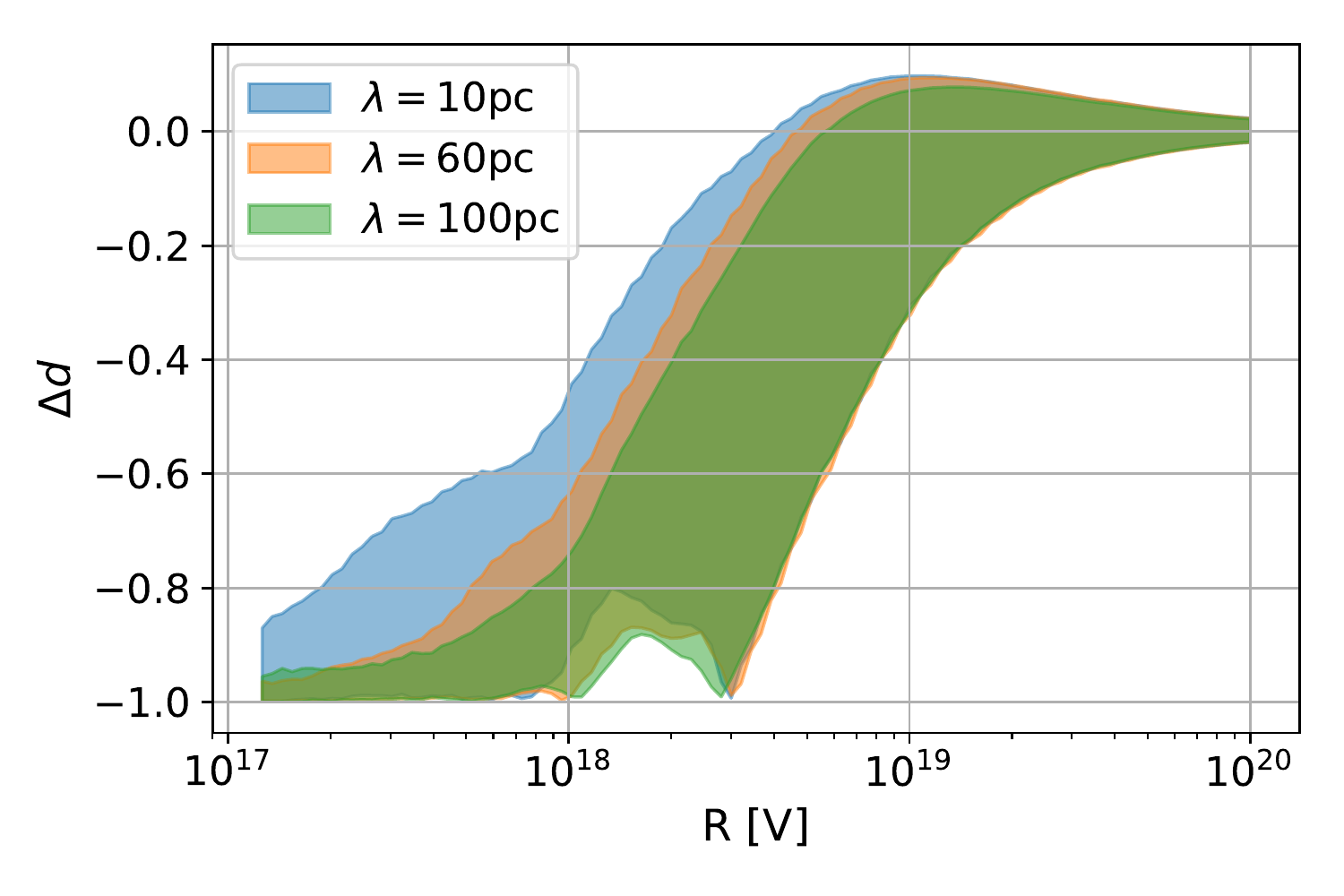}
\includegraphics[width=.31\linewidth]{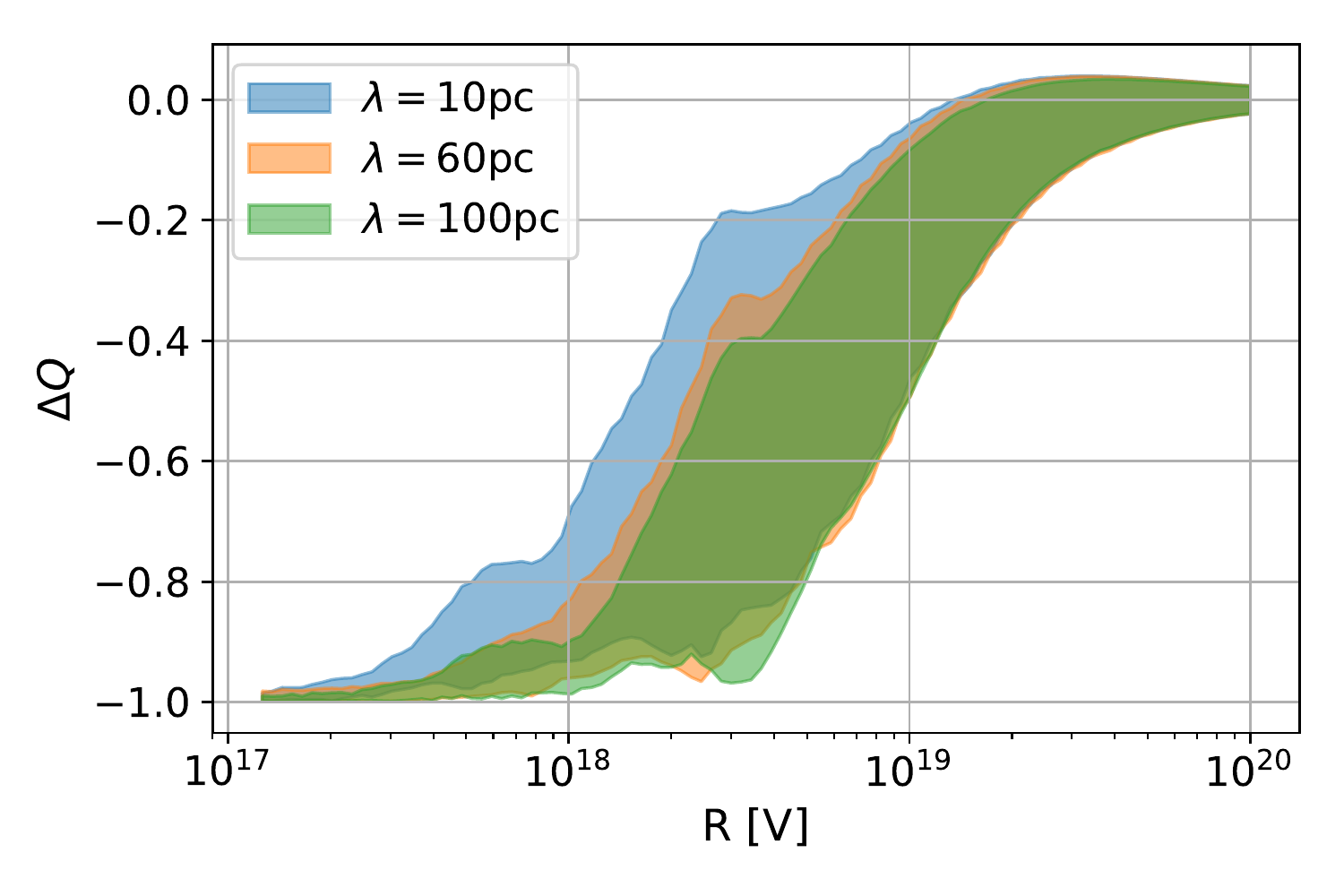}
\caption{Range of the modification of the anisotropy as a function of the
rigidity $R$. \textbf{Left:} Change of the dipole direction. \textbf{Middle:}
Change of the dipole amplitude. \textbf{Right:} Change of the quadrupole
amplitude.}
\label{modRange}
\end{figure}

\begin{figure}[htbp]
\centering
\includegraphics[width=.32\linewidth]{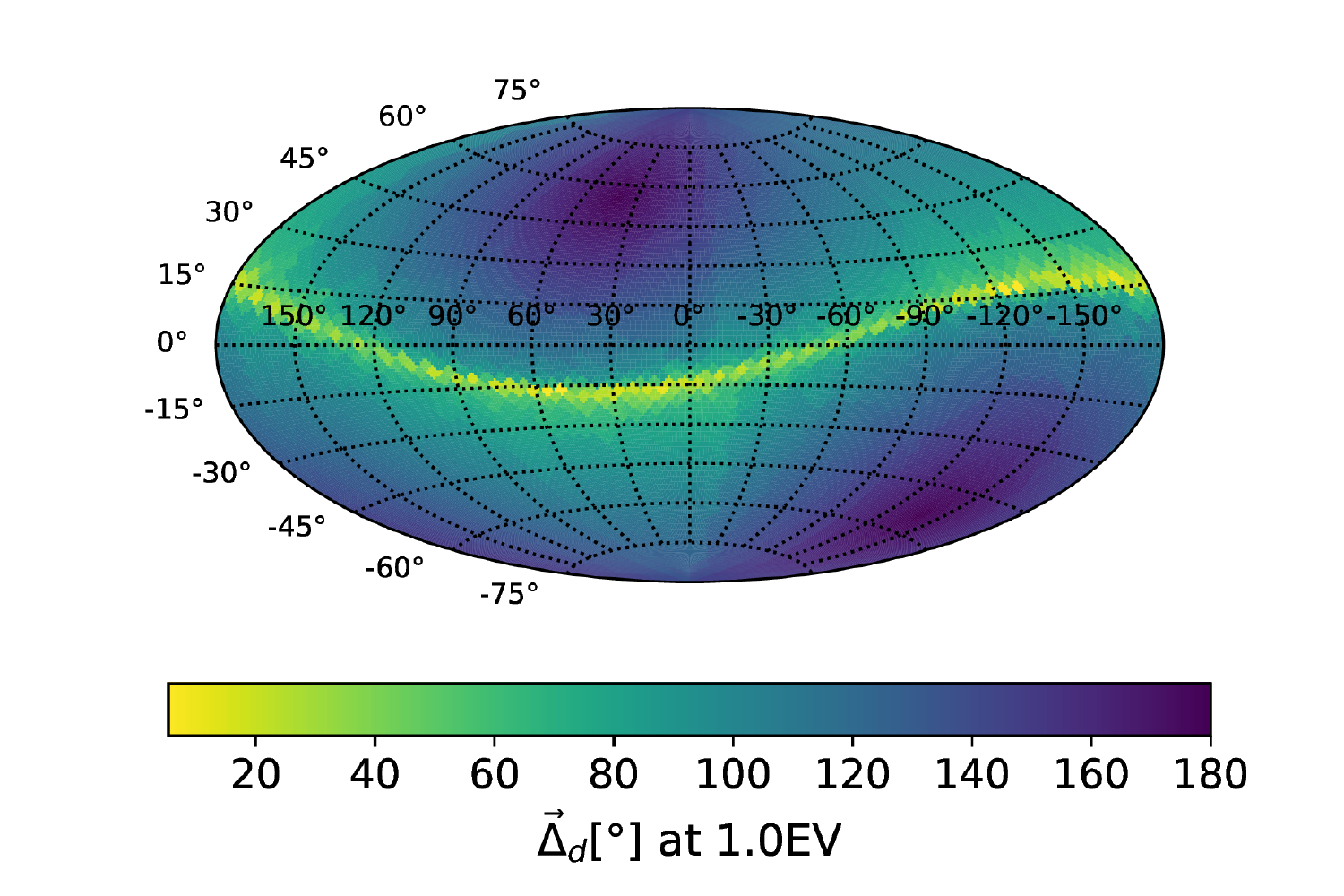}
\includegraphics[width=.32\linewidth]{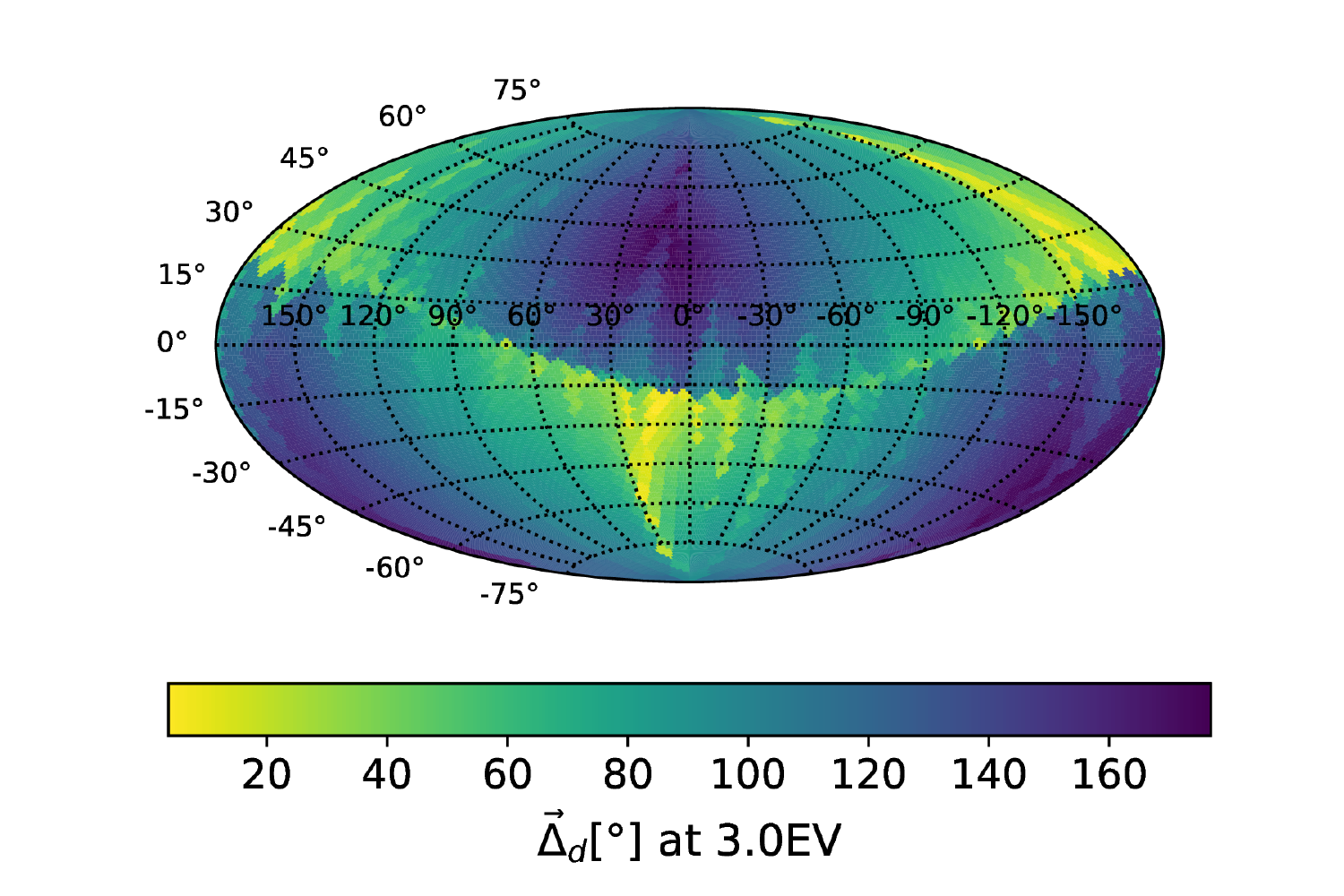}
\includegraphics[width=.32\linewidth]{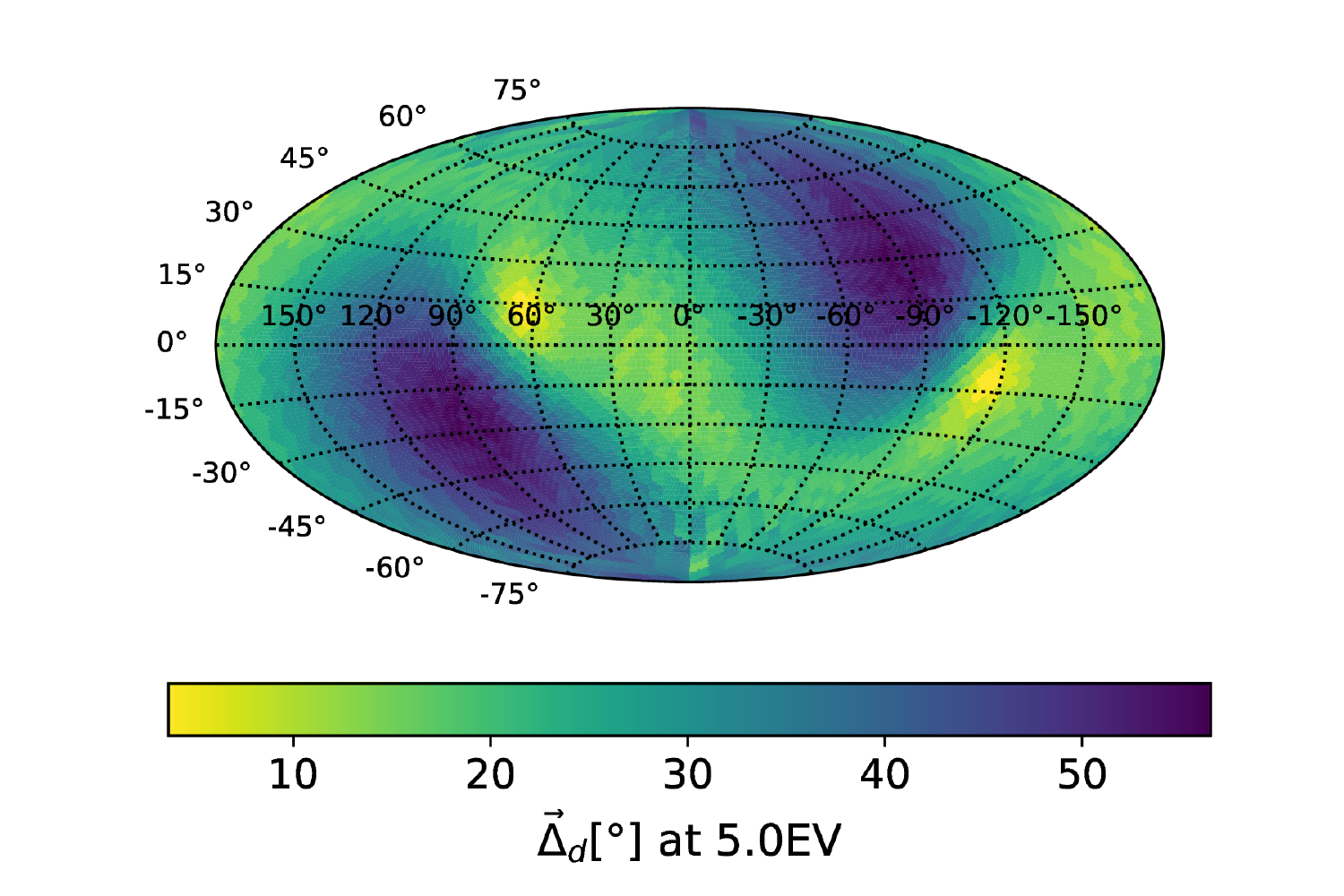}
\includegraphics[width=.32\linewidth]{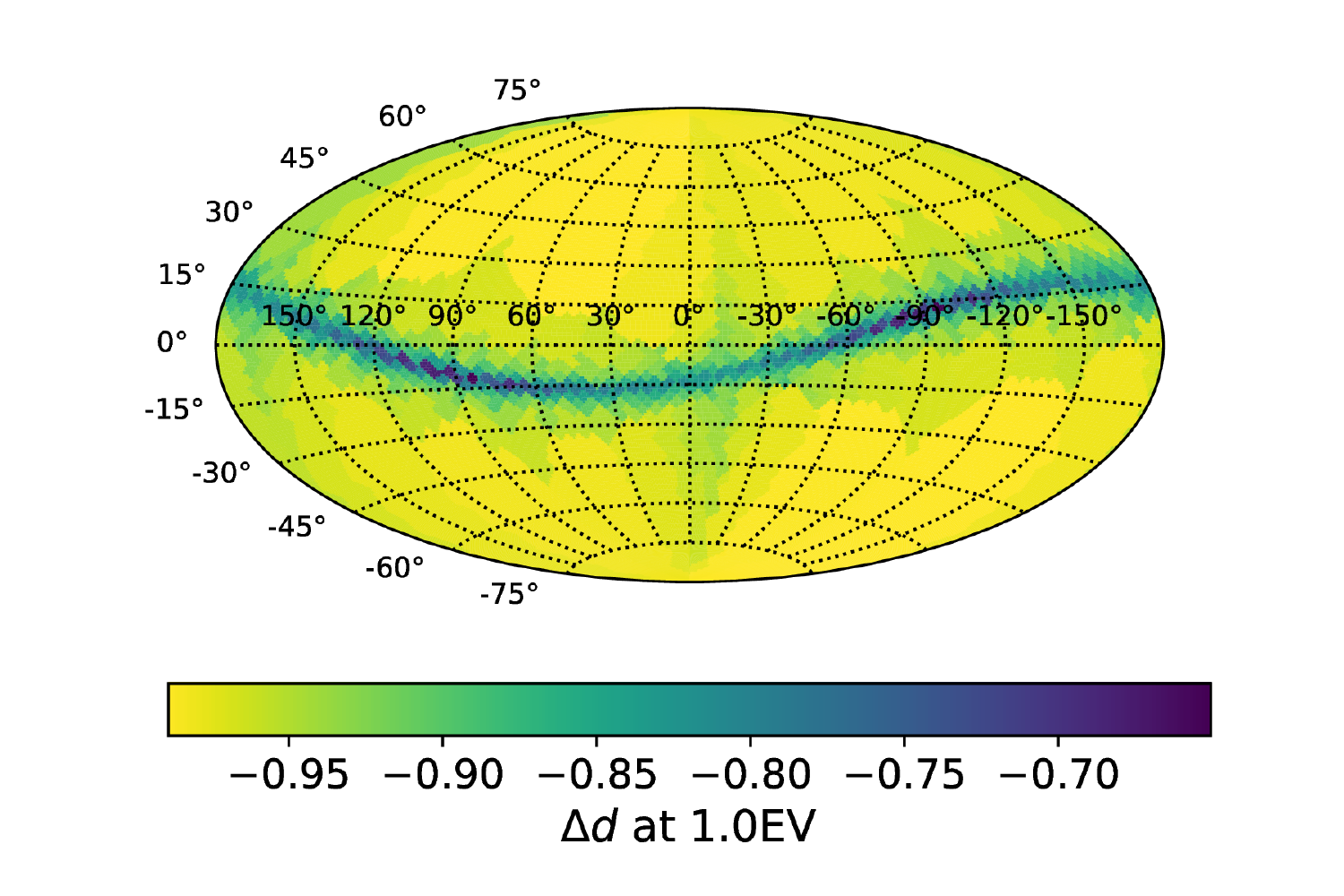}
\includegraphics[width=.32\linewidth]{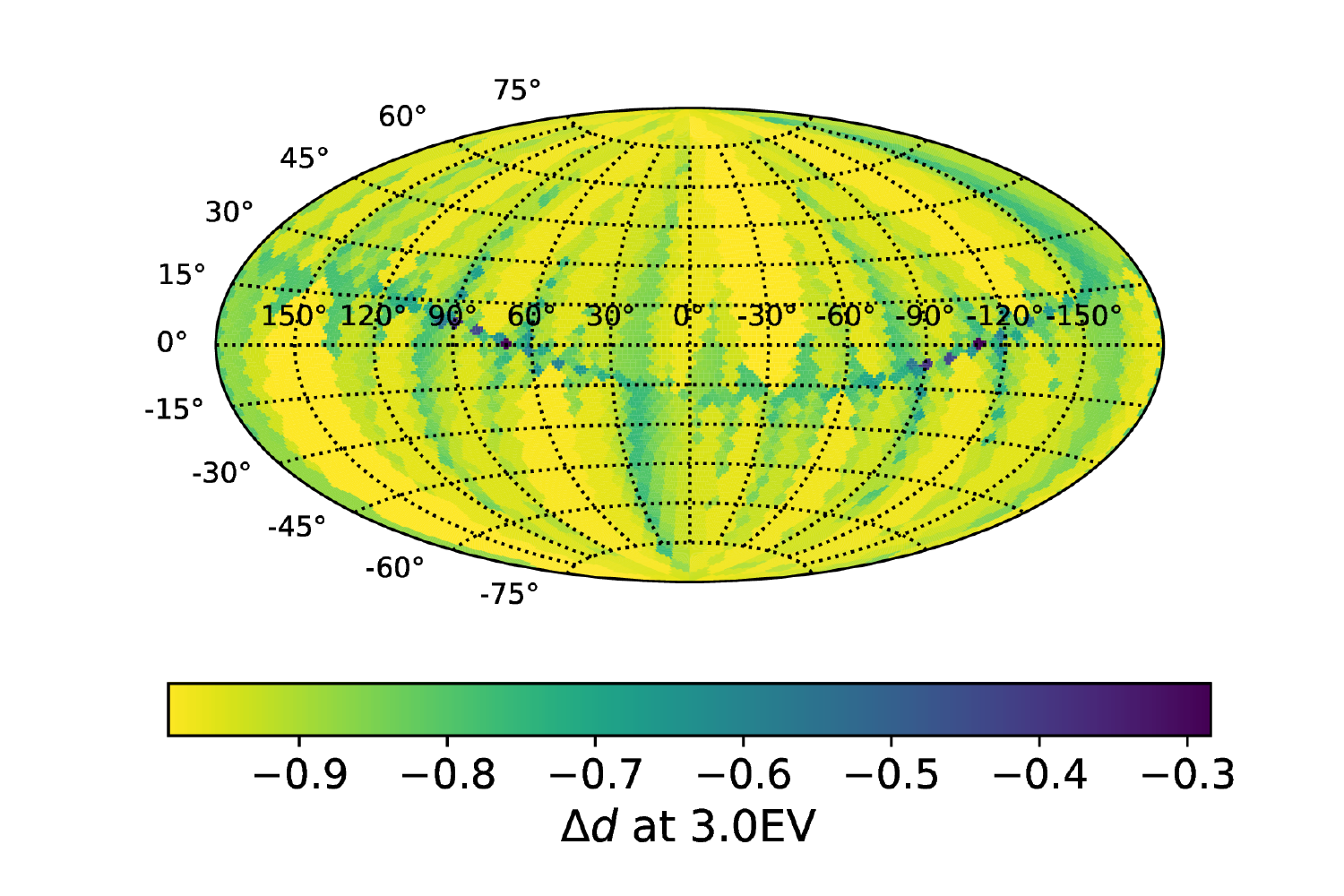}
\includegraphics[width=.32\linewidth]{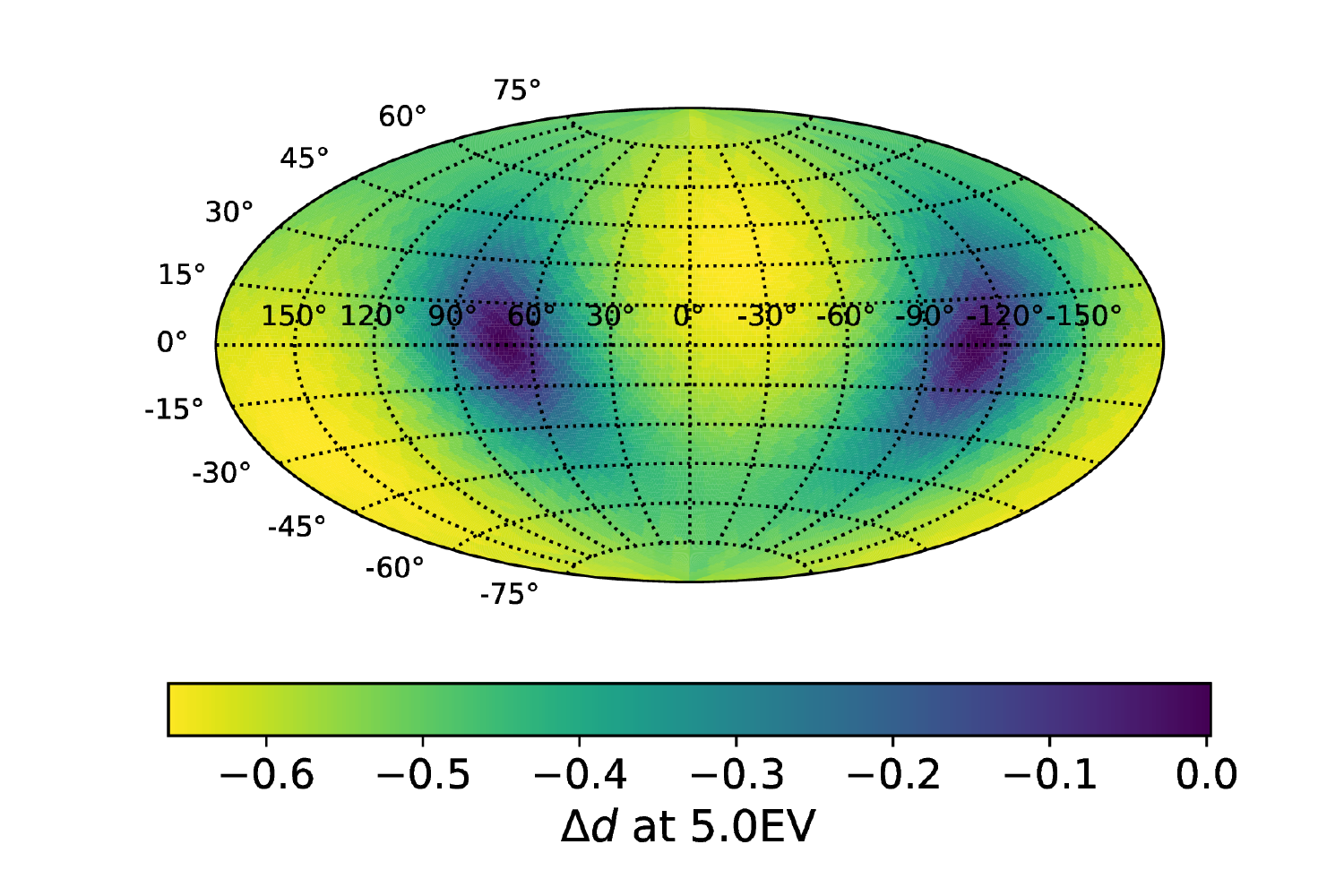}
\caption{Modification of the dipole direction (\textbf{upper panel}) and
amplitude (\textbf{lower panel}) dependent on its direction at Earth for three
different CR rigidities using $\lambda=60\,\text{pc}$. Here and hereafter, the
skyplots are displayed in Galactic coordinates.}
\label{modSky}
\end{figure}

The Galactic lens transforms the cosmic ray distribution from outside our Galaxy to
Earth, however, only the characteristics of the distribution at Earth are
accessible via experiments. Therefore, we first investigate the change
of the anisotropy --- direction and amplitude --- by the Galactic magnetic field. 
In general, the directions of the dipole are significantly shifted for
	some directions at rigidities $\lesssim 10\,\text{EV}$, as shown in the left
	Fig.~\ref{modRange}. According to the middle Fig.~\ref{modRange}, also the
	dipole amplitude decreases significantly at these rigidities, and thus, the
	Galactic magnetic field introduces anisotropies of higher order. In the case
	of the quadrupole, we obtain similar behaviour, but the amplitude is in
	principle reduced more by almost a factor two. The change of the dipole
	direction hardly dependents on the coherence length $\lambda$ of the
	turbulent magnetic field component, though in general the anisotropy amplitude decreases
	for an increasing $\lambda$. Hence, a change from $\lambda=10\,\text{pc}$ to
$\lambda=100\,\text{pc}$ can increase the suppression by about $20\,\%$ at a few EV.
At $R\lesssim 1\,\text{EV}$ the dipole completely vanishes for some directions
leading to a sudden decrease of the maximal value of $\vec\Delta_d$, as the
change of direction almost coincides with the change of amplitude --- see
Fig.~\ref{modSky}. Here, predominantly the narrow band of directions, that is
shown in the left Fig.~\ref{modSky}, still holds a non-vanishing dipole
providing a maximal change of direction of only about $90\degree$.

In the case of the dipole anisotropy, Fig.~\ref{modSky} show those
dipole directions at Earth, that are biased the most by the Galactic magnetic
field. At a few EeV the patterns expose a clear imprint by the Galactic
magnetic field leading to a narrow band of small dipole changes. Hereby, this
pattern exposes the symmetry of the used magnetic field model. At
$R=1\,\text{EV}$ a dipole that points away from this narrow band has almost
vanished, leading to some intriguing constraints on the maximal dipole
amplitude as discussed in the following. 

We can determine the anisotropy amplitudes outside our Galaxy
based on the observed amplitudes $d_{\rm obs}$ and $Q_{\rm obs}$, respectively, according to
\be
d_{\rm out} = \frac{d_{\rm obs}}{\Delta d+1}\,\text{ and } Q_{\rm out} = \frac{Q_{\rm obs}}{\Delta Q+1}\,.
\label{AmpOut}
\ee
Note, that the corresponding change of the anisotropy direction is only
taken into account for the dipole. 
Further, it needs to be satisfied that $d_{\rm out},\, Q_{\rm out}\leq 1$.
Therefore, we obtain a maximal dipole amplitude $\text{max}(d_{\rm
obs})=1+\Delta d$ at Earth dependent on the cosmic ray rigidity and the
direction of the observed dipole.  As indicated by the lower panel in
Fig.~\ref{modSky}, the suppression of the dipole amplitude at small rigidities
is huge. E.g.\ there is a huge range of directions of the dipole where its
observed amplitude at $R=1\,\text{EV}$ cannot exceed $\sim 1\,\%$. At
$R=5\,\text{EV}$ this range of directions has become smaller, but still there
are about two extended regions close to the Galactic center and its
anti-center, where the dipole amplitude cannot exceed $34\,\%$.

\subsection{Total flux bias for a dipole anisotropy}
Using the Eq.~\ref{AmpOut} we determine $(\Delta F)_{\rm d}$ for an ideal
observer as well as the difference $(\Delta F_{\rm Auger}-\Delta F_{\rm
TA})_{\rm d}$ between the different field of views of the observatories
dependent on the observed direction and amplitude of the dipole anisotropy.

\subsubsection{An ideal observer}
\begin{figure}[htbp]
\centering
\includegraphics[width=.31\linewidth]{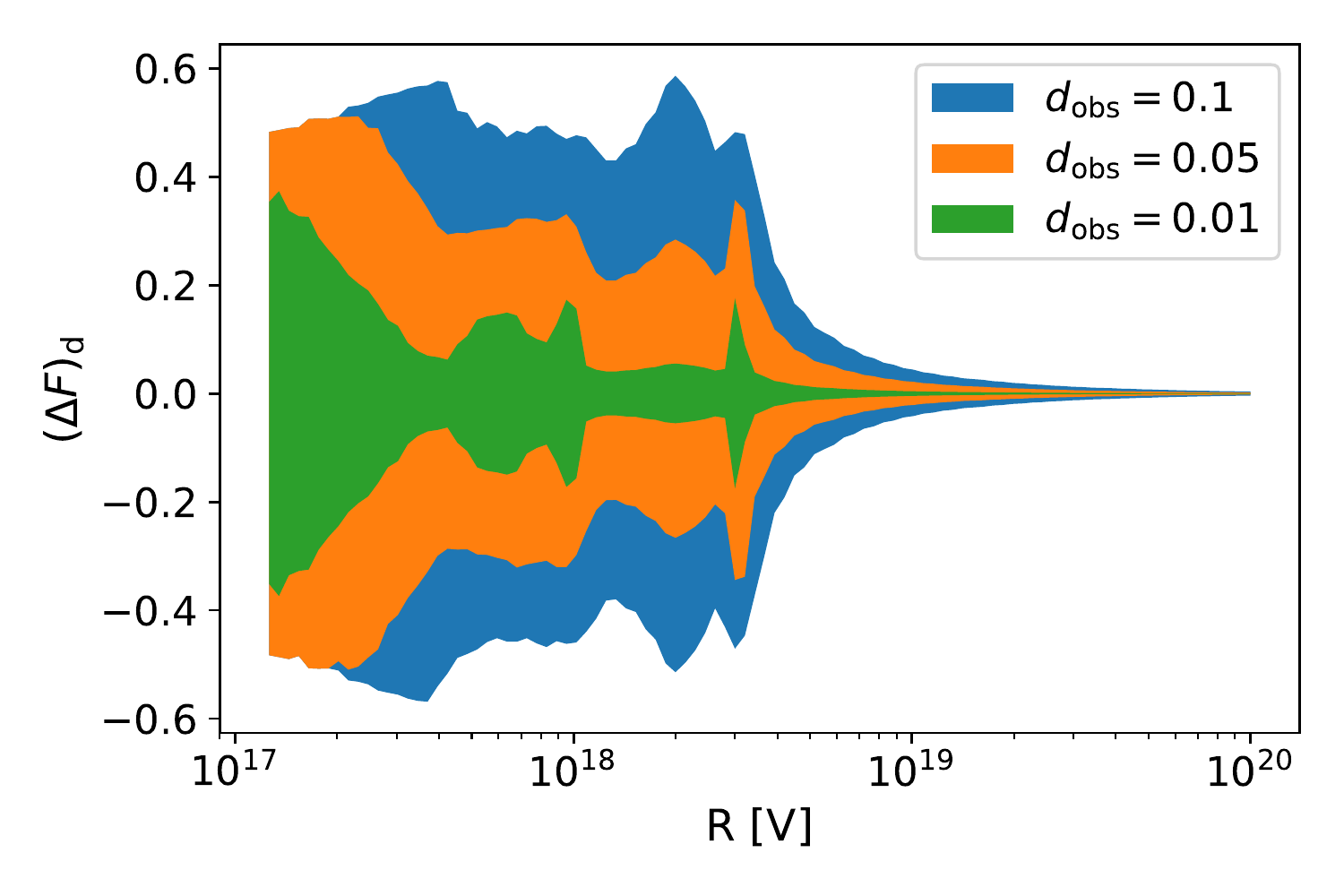}
\includegraphics[width=.32\linewidth]{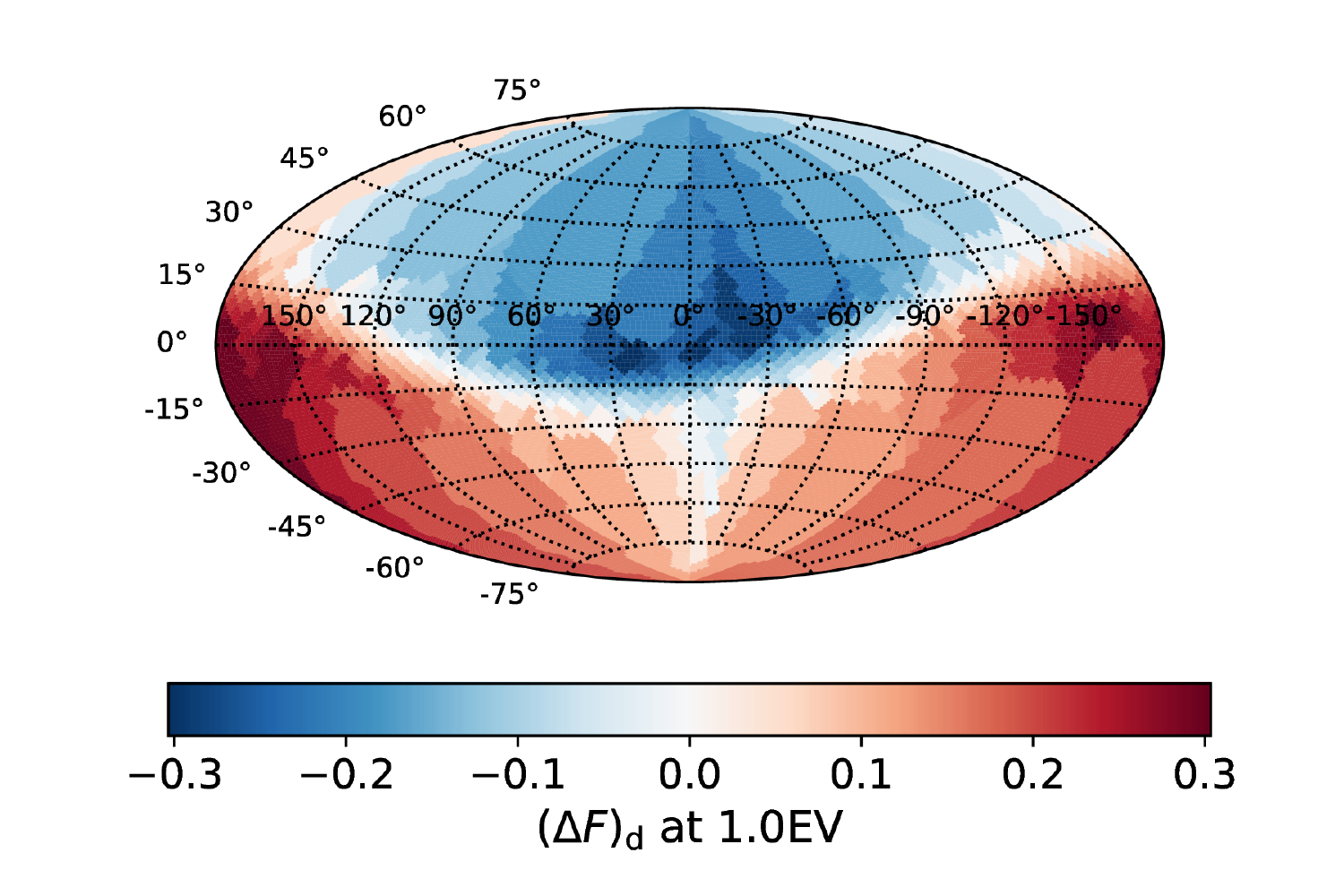}
\includegraphics[width=.32\linewidth]{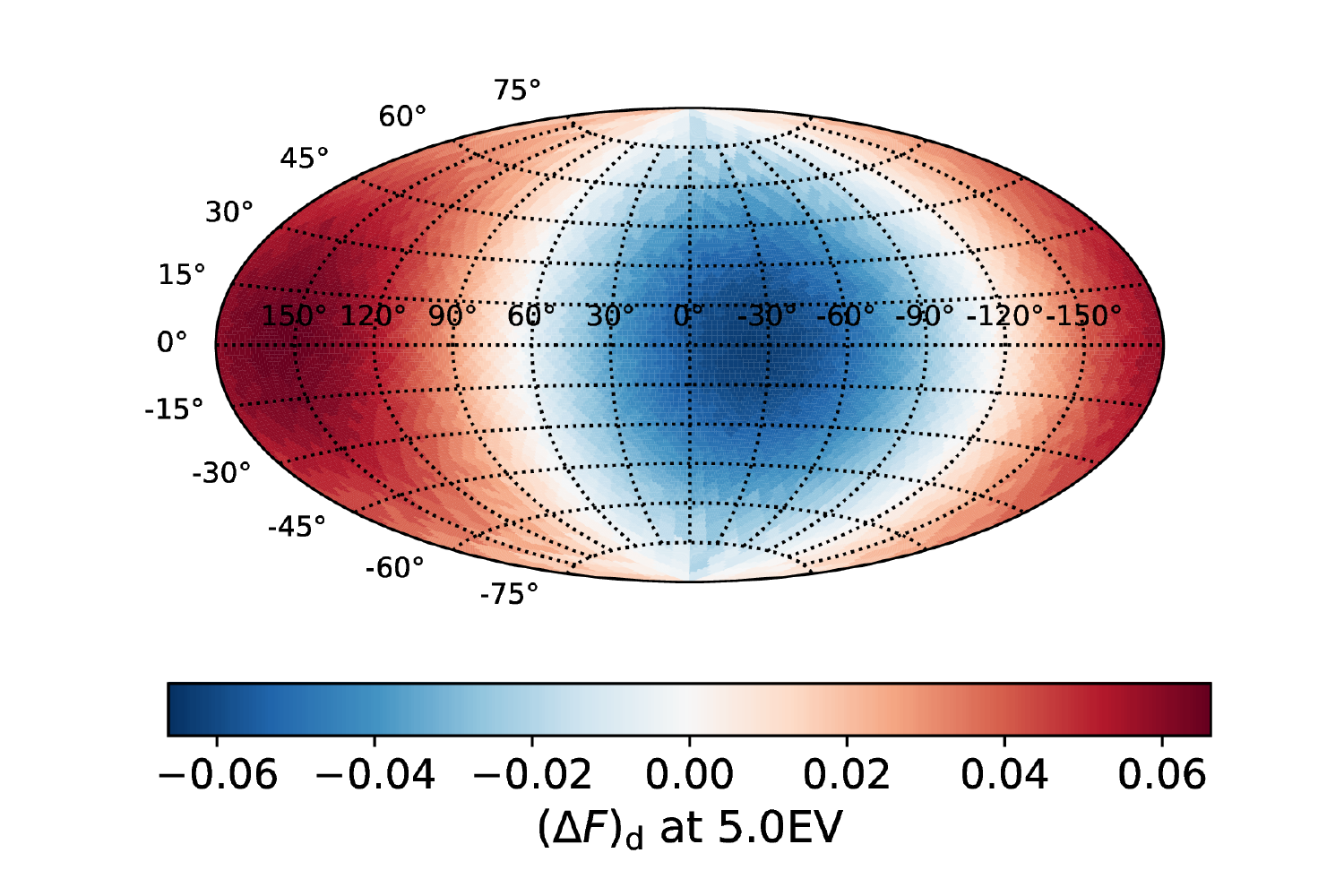}
\caption{Modification of the UHECR flux due to a dipole anisotropy for an ideal
observer. \textbf{Left:} Dependent on the rigidity. \textbf{Middle (right):}
Dependent on the dipole direction using a CR rigidity of $R=1\,\text{EV}$
($R=5\,\text{EV}$) for an amplitude $d_{\rm obs}=0.05$ at Earth.}
\label{modDip}
\end{figure}

Here, the left Fig.~\ref{modDip} shows the band of $(\Delta F)_{\rm d}$, as the
certain value depends on the direction of the dipole, in the case of a dipole
anisotropy with three different amplitudes at Earth. At $R< 10\,\text{EV}$ the
bands widen significantly and at certain rigidities we obtain a huge change of
$(\Delta F)_{\rm d}$ for certain dipole directions. The skyplots in the middle
and right Fig.~\ref{modDip} reveal that these directions are about to correlate
to the Galactic center and its anti-center. At small rigidities the pattern is
almost symmetric in longitude, whereas at higher rigidities it becomes
symmetric with respect to the galactic latitude.

Note that the unexpected increase of $(\Delta F)_{\rm d}$ in the left
Fig.~\ref{modDip} at small rigidities for a strong dipole amplitude $d_{\rm
obs}$ is only an artifact of the constrain $d_{\rm out}\leq 1$. Thus, at these
rigidities the dipole directions of the maximal modification coincide with
those where $\text{max}(d_{\rm obs})$ is smaller than the requested amplitude
$d_{\rm obs}$. In these cases we change $d_{\rm obs}$ to $\text{max}(d_{\rm
obs})$, so that $d_{\rm out}=1$.

\subsubsection{Difference between Auger and TA}
\begin{figure}[htbp]
\centering
\includegraphics[width=.31\linewidth]{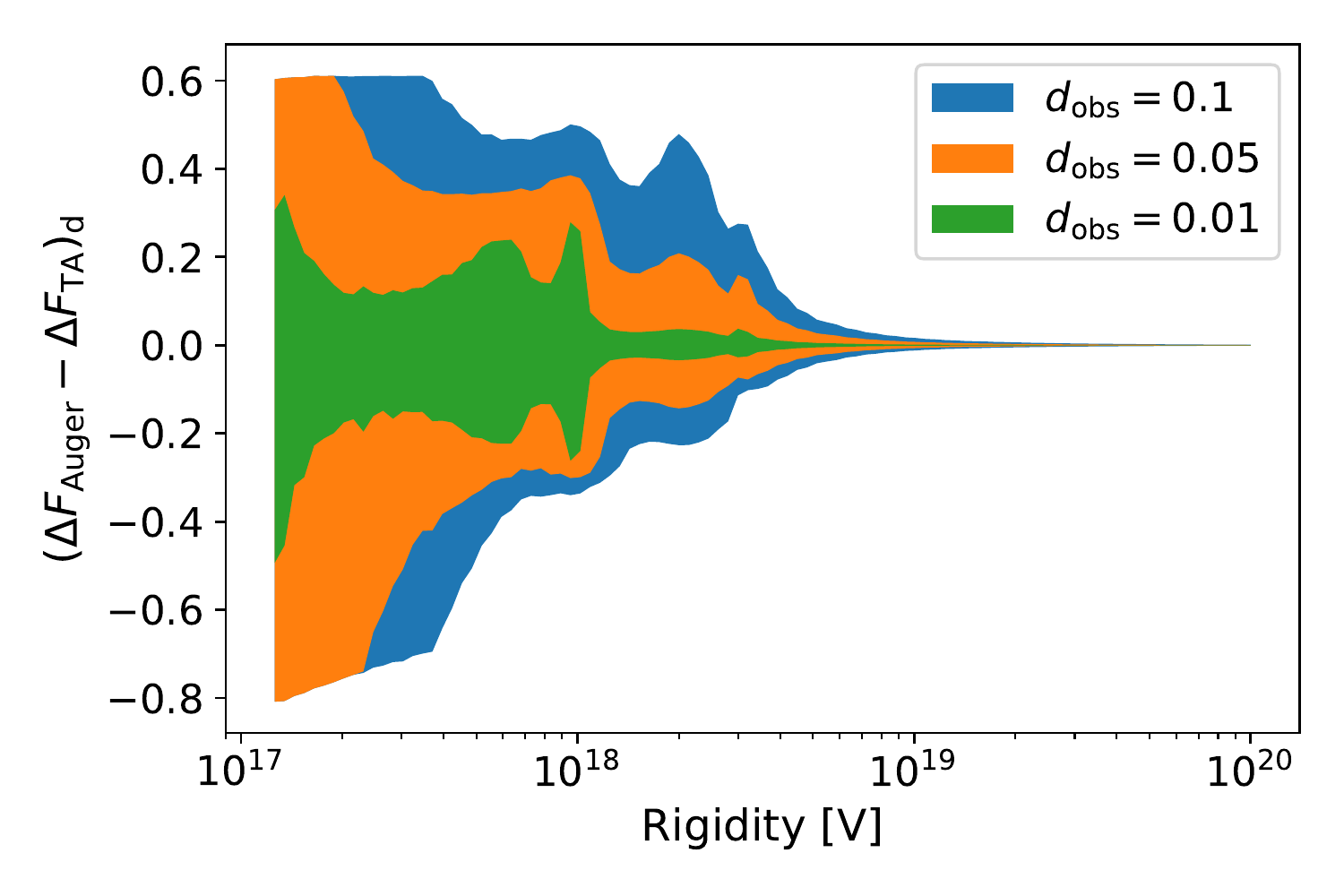}
\includegraphics[width=.32\linewidth]{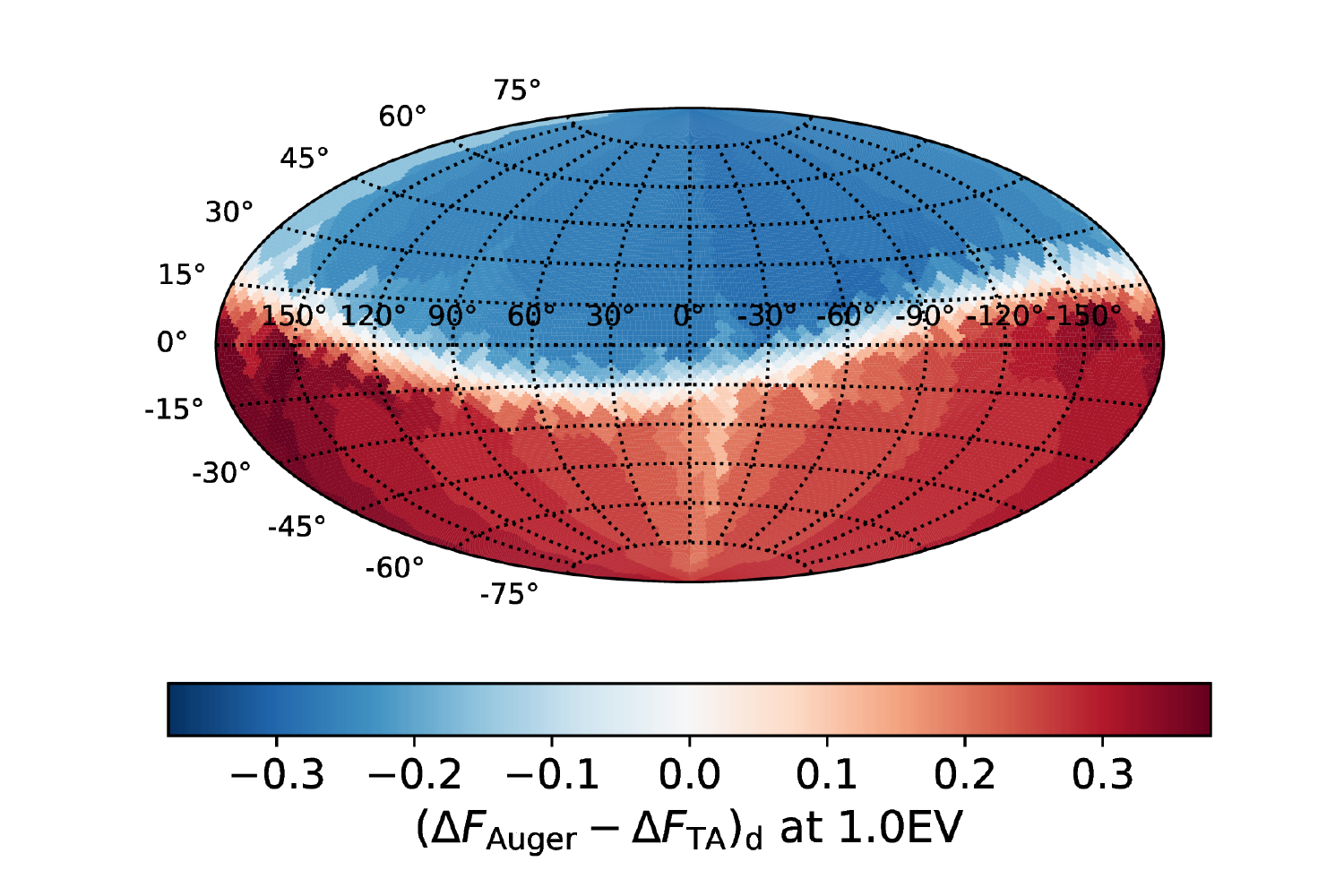}
\includegraphics[width=.32\linewidth]{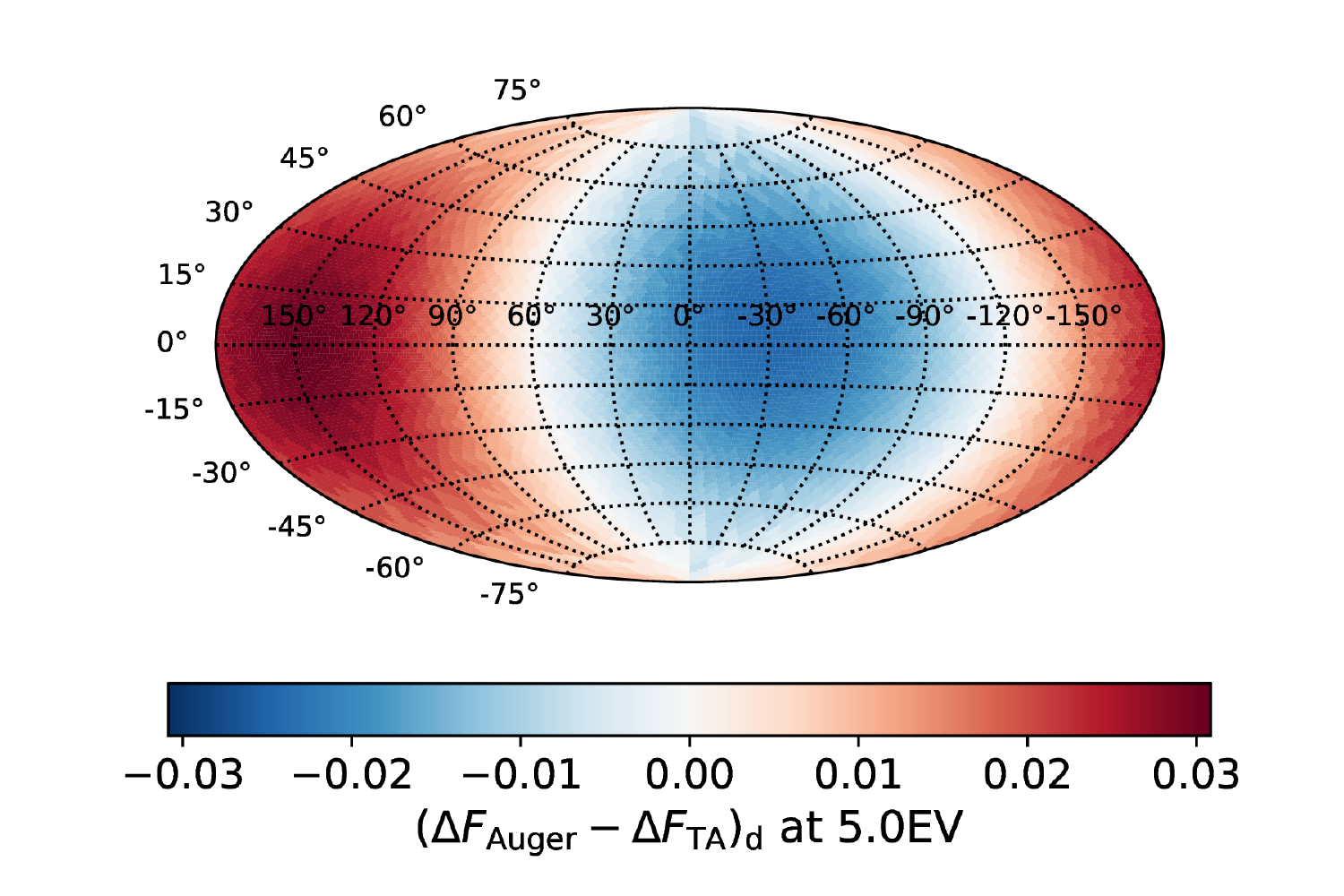}
\caption{Difference of the UHECR flux due to a dipole anisotropy in the field
of view of Auger and TA. The plot parameters and characteristics are the same
as given in Fig.~\ref{modDip}.}
\label{modDip2}
\end{figure}

In the following, we take the different field of views of the experiments into
account and determine the flux difference between them. So, the bands of
$(\Delta F_{\rm Auger}-\Delta F_{\rm TA})_{\rm d}$ in the left
Fig.~\ref{modDip2} shows in principle a similar rigidity dependence as $(\Delta
F)_{\rm d}$ in the left Fig.~\ref{modDip}. At $R\leq 1\,\text{EV}$ the bands
for $d_{\rm obs}\leq 0.05$ widen significantly with decreasing rigidity, so
that even for a very small dipole amplitude a difference of more than $15\%$
becomes possible for certain dipole directions. The directional pattern, as
shown the middle and right Fig.~\ref{modDip2}, does not show any obvious
correlation to the different field of views. Instead they reveal a similar
directional pattern as in Fig.~\ref{modDip}, hence, the directional dependence
is less influenced by the different field of views than by the Galactic
magnetic field.

In total, the Galactic magnetic field can lead to differences of more than
$10\%$ in the total flux that is observed by Auger and TA. However, the exact
value strongly depends on the directions of the anisotropy, and the rigidity of
the particle.  To draw a conclusive answer on the UHECR flux modification by
the Galactic magnetic field, we need to include the observed chemical
composition as well as the direction of the dipole.

\subsection{Total flux bias for a quadrupole anisotropy}
\begin{figure}[htbp]
\centering
\includegraphics[width=.41\linewidth]{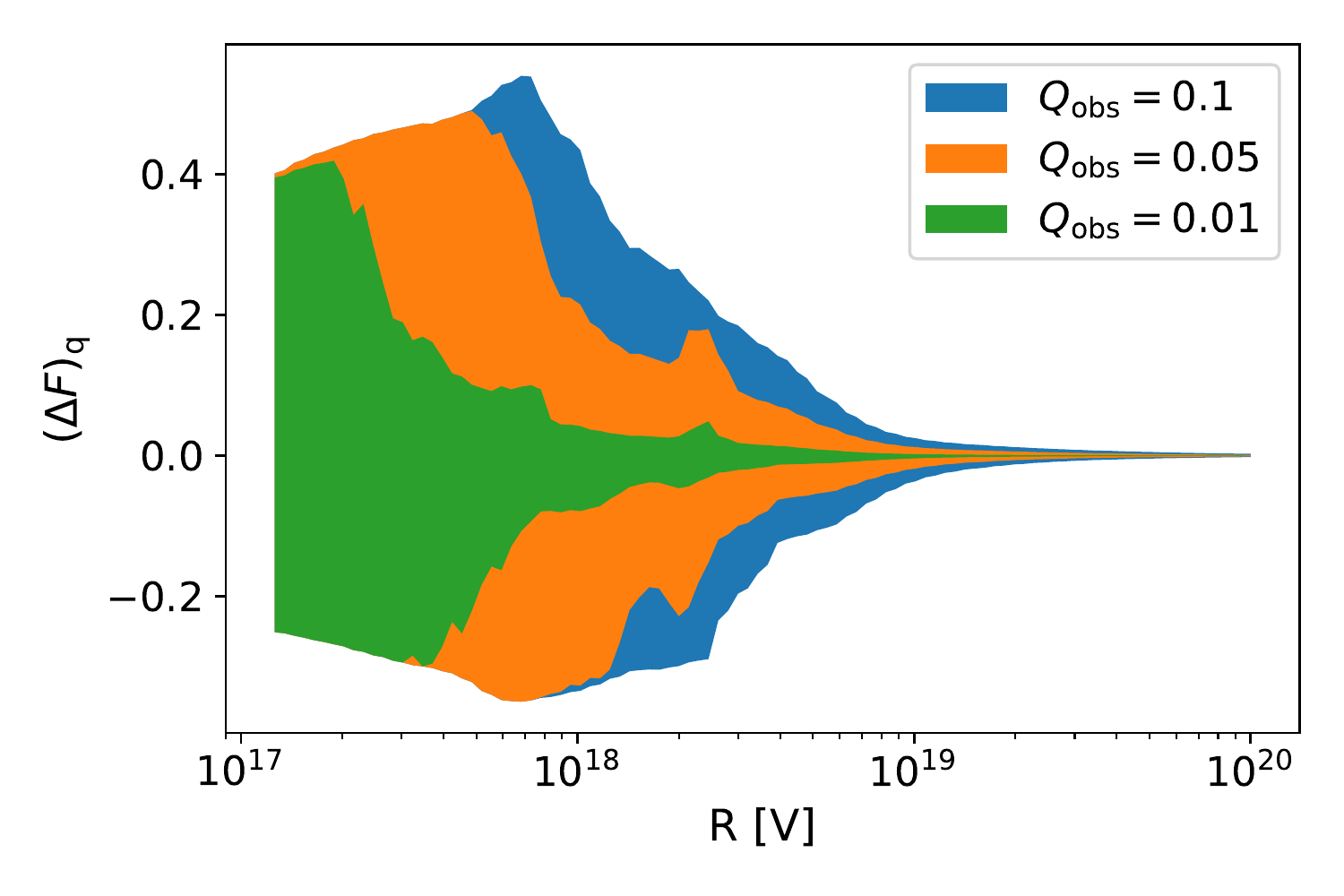}
\includegraphics[width=.41\linewidth]{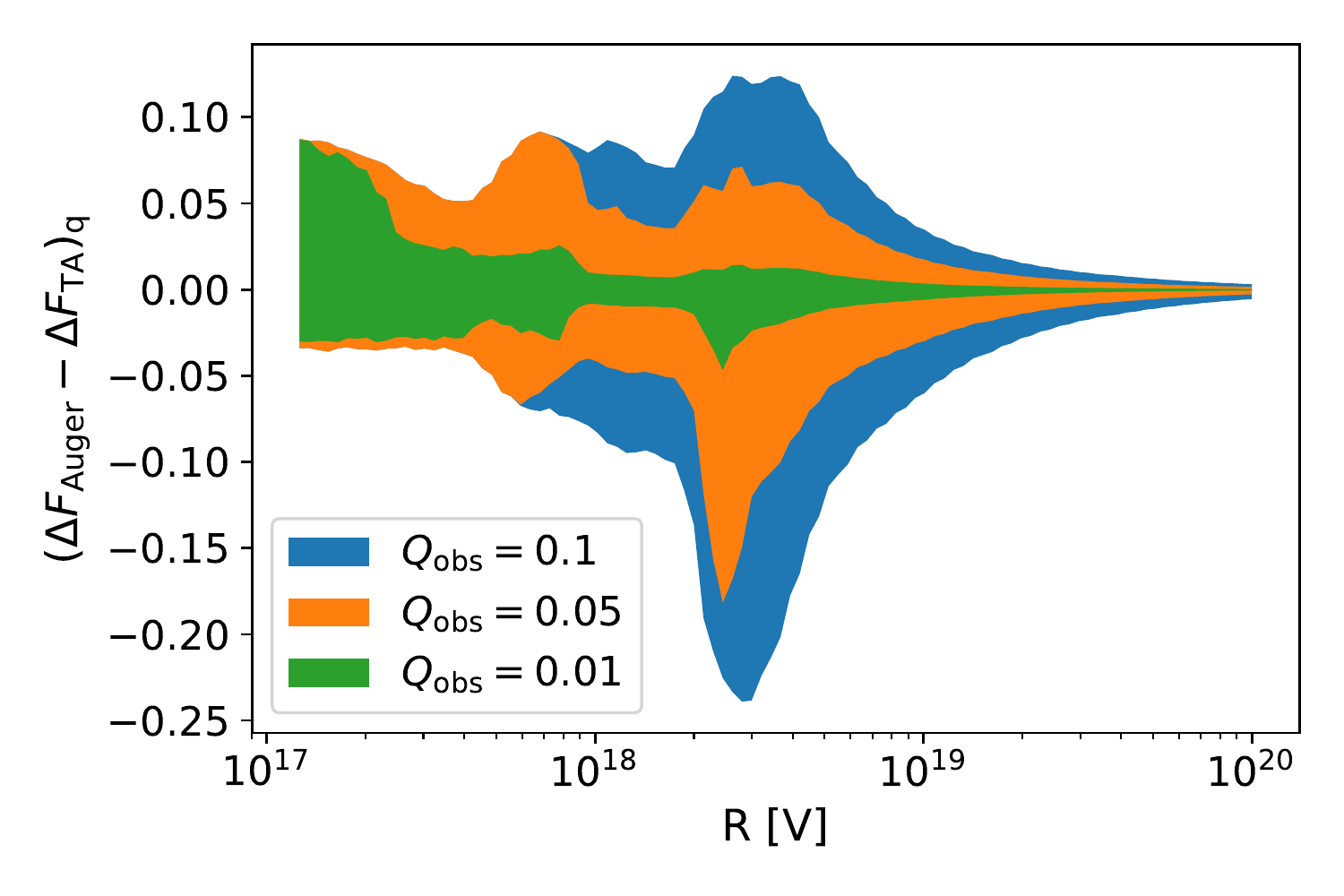}
\caption{UHECR flux bias due to a quadrupole anisotropy. \textbf{Left:} For an
ideal observer. \textbf{Right:} Difference between the field of view of Auger
and TA.}
\label{modQuad}
\end{figure}

In the case of the quadrupole anisotropy, we are unable to correlate the
observed direction of the quadrupole to the one outside the Galaxy.
Still we can provide the range of the bias based on the range of $\Delta Q$, as
given in the right Fig.~\ref{modRange}.
The left Fig.~\ref{modQuad} shows that $(\Delta F)_{\rm q}$ in the case of a
symmetric quadrupole anisotropy yields a similar pattern as in the case of a
dipole. However, the bands are no longer symmetric indicating that there are
quadrupole directions that yield to a stronger amplification than suppression
of the flux. Although, at certain rigidities, the suppression can still
dominate.  Further, there the bands show less peaks and the constrain $Q_{\rm
out}\leq 1$ applies even for $Q_{\rm obs}=0.01$ at small rigidities. On average
the quadrupole anisotropy yields $(\Delta F)_{\rm q}$ values that are a bit
smaller than the $(\Delta F)_{\rm d}$ values.

Further, the right Fig.~\ref{modQuad} shows
the corresponding flux difference $(\Delta F_{\rm Auger}-\Delta F_{\rm
TA})_{\rm q}$ between the field of view of Auger and TA.  Interestingly, there
is no general decrease of the magnitude of the differences between the two
experiments, but the difference reaches a maximum at about \SI{3}{\exa\volt}.

\section{Bias based on the Auger data}
\label{results2}
The relevance of the bias by the Galactic magnetic field considerably depends
on the cosmic ray's rigidity and the direction of the anisotropy, as well as the
magnetic field itself. Here, we use the JF12 model again with a coherence
length of $\lambda=60\,\text{pc}$, unless otherwise stated.
Further, we use the observed mean logarithm of the
mass number $\langle \ln A \rangle$ \cite{PhysRevD.90.122005}, and the
information on the dipole strength and direction \cite{Aab_2018} in the
following to provide a constrain on this effect for the observed dipole
anisotropy.
\subsection{Total flux in case of the Auger dipole}
The particle's rigidity $R_{\rm obs}$ is approximated by
\begin{equation}
    R_{\rm obs} = \begin{cases} E_{\rm obs}/(A/2), \text{ for }A\geq 2\,\\
    E_{\rm obs}, \text{ for }A< 2\,,
    \end{cases}
\end{equation}
where $E_{\rm obs}$ denotes the observed median energies of the dipole, as
given in Table \ref{DipParameters}, and the mass number $A=\exp(\langle \ln A
\rangle)$. Further, we use the derived power-law behavior of the dipole
amplitude \cite{Aab_2018}
\begin{equation}
    d_{\rm obs} = 0.055\,\left( E_{\rm obs}/10\,\text{EeV}\right)^{0.79}
    \label{obsDipoleAmpl}
\end{equation}
and the observed directions $(l, b)_{\rm obs}$ of the dipole --- see Table
\ref{DipParameters}. At energies that are not covered by the data of $\langle
\ln A \rangle$ or $(l, b)_{\rm obs}$ we use linear interpolation, where the
boundary values are used outside the observed energy range.  Since these data
values can in principle change from outside our Galaxy to Earth, we use $\Delta
d$ dependent on $(l, b)_{\rm obs}$ and $R_{\rm obs}$ in order to estimate the
dipole outside our Galaxy dependent on the hadronic interaction model as listed
in Table \ref{DipParameters}. Hereby, we also include the shift of the
dipole direction due to the Galactic magnetic field. We suppose that neither
the energy of UHECRs nor their chemical composition changes during the
propagation through the Galaxy, which this is a necessary condition to apply
the backtracking approach in the first place.

\begin{table*}[t]
\centering
\footnotesize
\caption{Parameters of the Auger dipole.}
  \begin{tabular}{c c c c c c c c c}
  \toprule
             $E_{\rm obs}\,[\text{EeV}]$ & $d_{\rm obs}$ & $(l,\,b)_{\rm obs}\,[\degree]$ & $d_{\rm out,1}$  &
$d_{\rm out,2}$  & $d_{\rm out,3}$ & $(l,\,b)_{\rm out,1}\,[\degree]$ & $(l,\,b)_{\rm out,2}\,[\degree]$ & $(l,\,b)_{\rm out,3}\,[\degree]$ \\ 
  \midrule
    $5$ & $0.032$ & $(-73,\,-32)$ & $0.183$ & $0.057$ & $0.051$ & $(3,\,0)$ & $(-53,\,-57)$ & $(-45,\,-51)$    \\ 
    $10.3$ & $0.056$ & $(-139,\,-3)$ & $0.159$ & $0.074$ & $0.067$ & $(-155,\,14)$  & $(-141,\,7)$  & $(-138,\,5)$    \\
    $20.2$ & $0.096$ & $(-103,\,-34)$ & $0.124$ & $0.099$ & $0.097$ & $(-111,\,-45)$  & $(-93,\,-42)$  & $(-96,\,-39)$    \\
    $39.5$ & $0.163$ & $(-101,\,-11)$ & $0.157$ & $0.151$ & $0.151$ & $(-79,\,-17)$  & $(-87,\,-14)$  & $(-87,\,-14)$    \\
   \bottomrule
   \multicolumn{9}{l}{\footnotesize Here, $d_{{\rm out},i}$ and $(l,\,b)_{{\rm out},i}$ refer to the resulting dipole features outside our Galaxy using EPOS-LHC ($i=1$),} \\
   \multicolumn{9}{l}{\footnotesize Sibyll$2.1$ ($i=2$), and QGSJetII-04 ($i=3$). }
\end{tabular}
  \label{DipParameters}
\end{table*}

\begin{figure}[htbp]
\centering
\includegraphics[width=.32\linewidth]{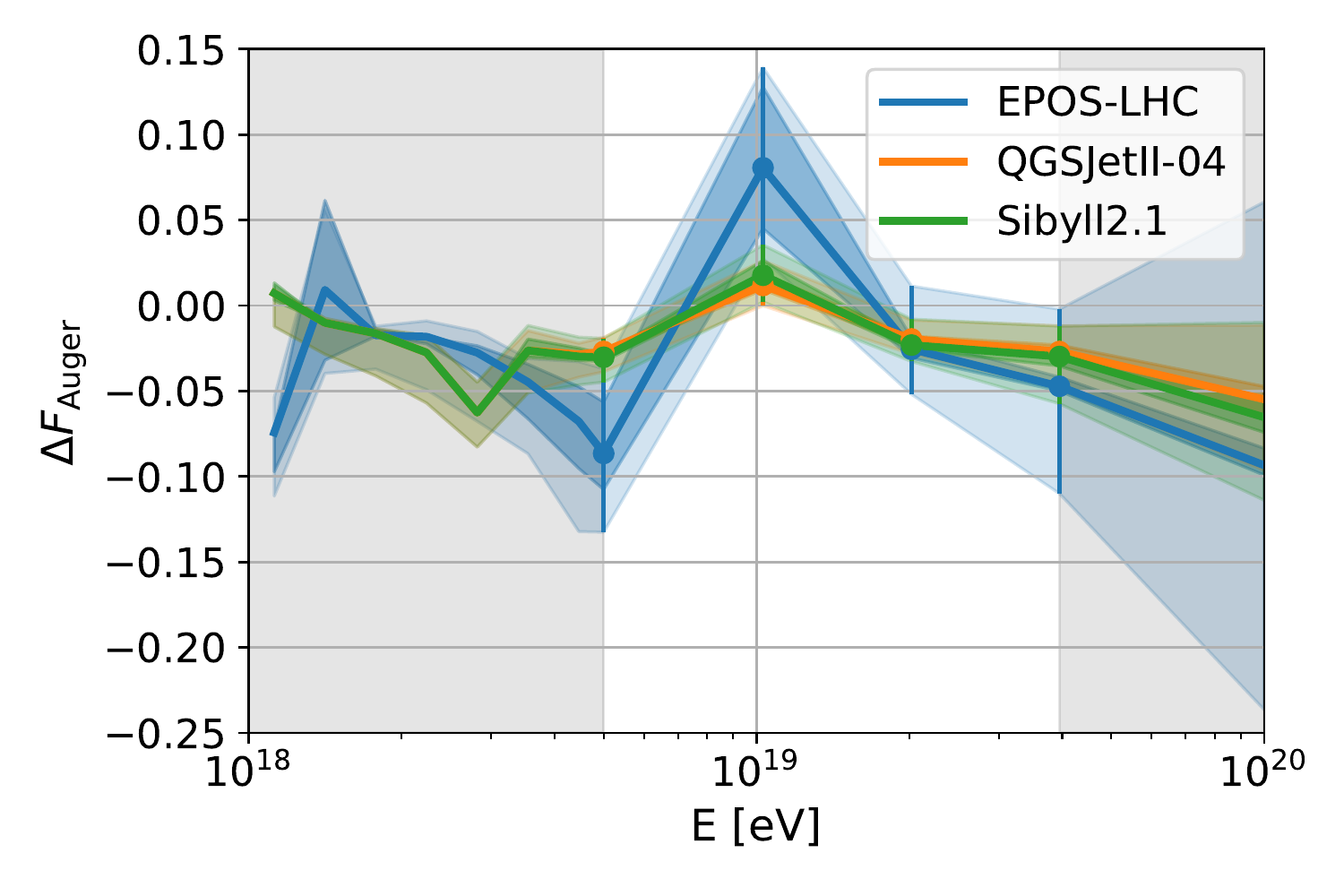}
\includegraphics[width=.32\linewidth]{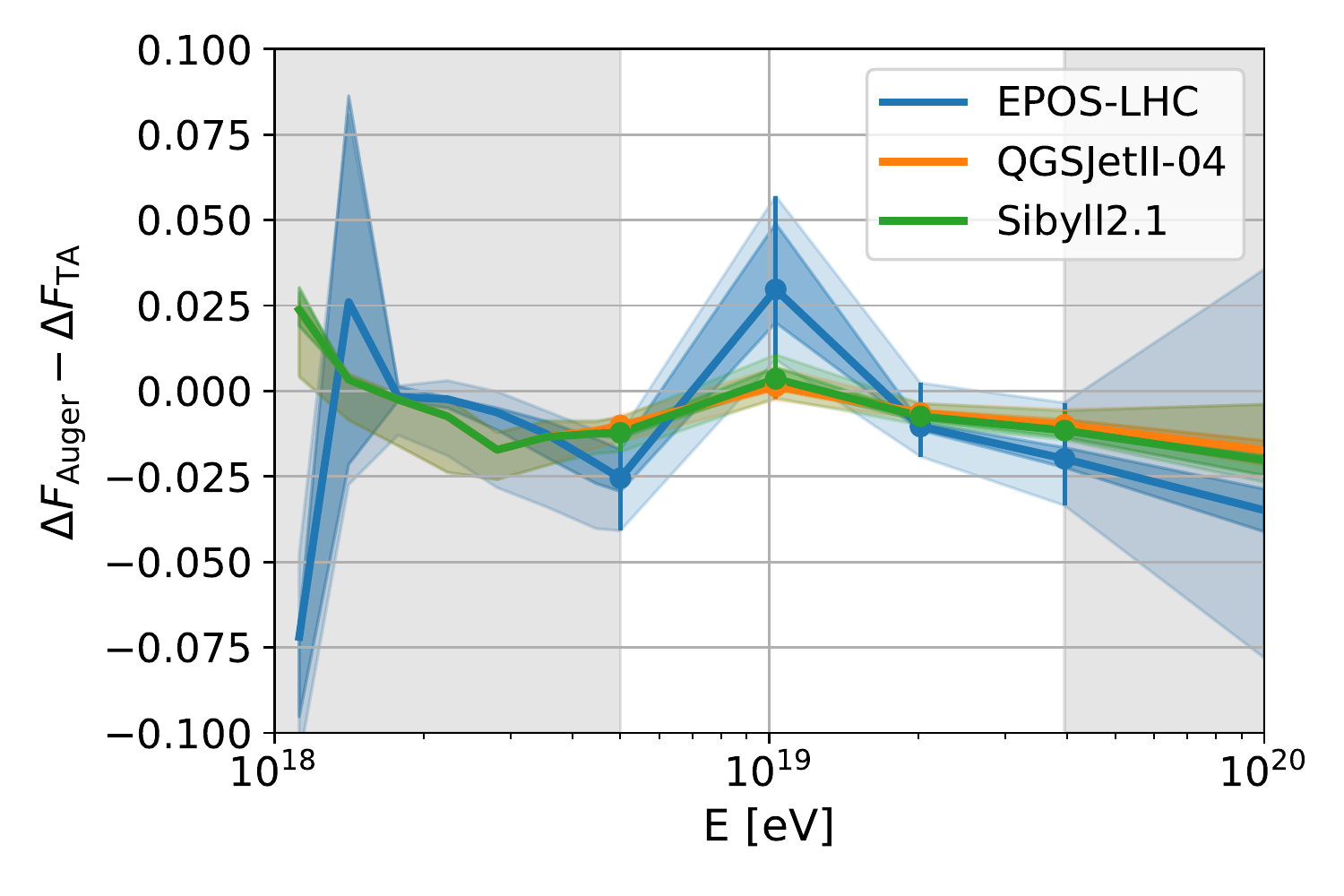}
\includegraphics[width=.32\linewidth]{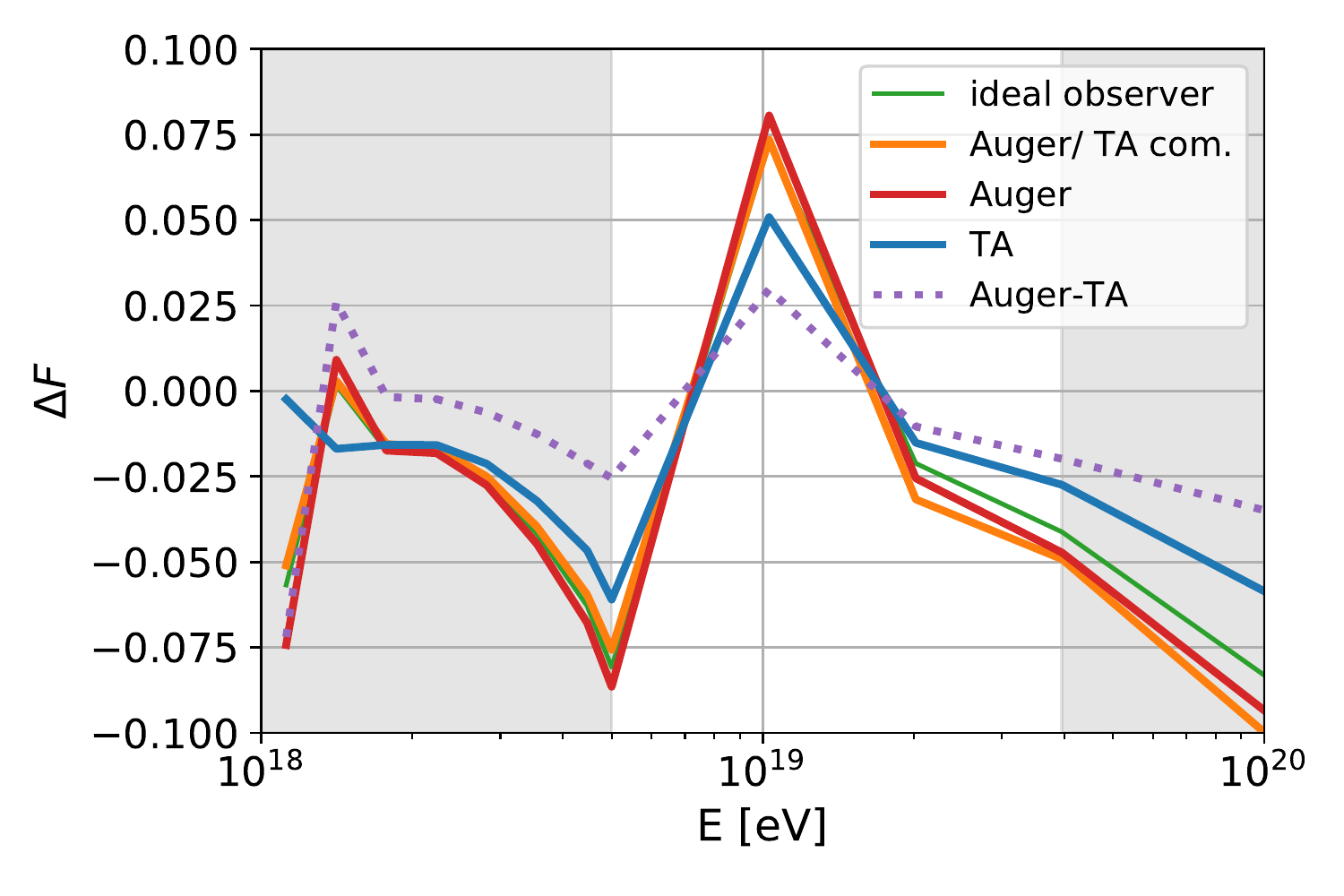}
\caption{The UHECR flux bias and difference based on the Auger data using a
	coherence length of $\lambda=60\,\text{pc}$. Due to the lack of observational
	data, the dipole direction is not changed for $E<5\,\text{EeV}$ and
	$E>39.5\,\text{EeV}$, which is indicated by the shaded background.
	\textbf{Left (middle):} The bias in Auger's field of view (the difference
	between the field of view of Auger and TA) for three different hadronic
	interaction models. The statistical uncertainties of $\langle \ln A \rangle$
	refers to the darker error band and the total error which includes the
	directional uncertainty of the dipole refers to the lighter band.
	\textbf{Right:} Using the hadronic interaction model 'EPOS-LHC' and the mean
	dipole direction, the bias is displayed for a full sky observatory (green
line), Auger (red line), TA (blue line), the common observation band of Auger
and TA (orange line), as well as the difference between Auger and TA (dotted
purple line).} \label{modObsData}
\end{figure}

As shown in Fig.~\ref{modObsData}, the UHECR flux is reduced at energies $\ll 10\,\text{EeV}$ by
several percentage --- the exact value strongly depends on the hadronic
interaction model --- in the case of the observed dipole. Though, at about
$10\,\text{EeV}$ an amplification of at most $14\%$ is obtained. As expected,
the hadronic interaction model 'EPOS-LHC' \cite{PhysRevLett.101.171101,
PhysRevC.92.034906} that predicts the most heavy composition also leads to the
largest values of $\Delta F$. Further, the statistical uncertainties of
	$\langle \ln A \rangle$ as well as the directional uncertainty of the dipole
	direction lead to a wide range of $\Delta F$ values at a few EeV, in
	particular for the 'EPOS-LHC' model. Hence. at about $100\,\text{EeV}$, the
	flux can be amplified by 5\,\% or suppressed by more than 20\%.
Comparing this with the statistical uncertainty of the flux measurement of the
Pierre Auger Observatory, which is between 0.5\% and  5\% below the cut-off at
approx.~\SI{30}{\EeV}, all model predictions lead to a significant total flux
bias at about 3 and \SI{5}{\EeV}, respectively, at least. 

The right Fig.~\ref{modObsData} exposes that the previously described
	trend applies not only for Auger's field of view but also for the common
	observation band of Auger and TA, i.e.\ $-15.7\degree<\delta<24.8\degree$, as
	well as the ideal observer. In all cases the flux is amplified at about
	$10\,\text{EeV}$, but suppressed below and above this energy yielding a
	significant change of the spectral behavior in this energy range. At energies
	below some tens of EeV the observed flux is well described by a series of
	broken power laws \cite{2019ICRC...450}, hence, $F_{\rm obs}(E)\propto
	E^{-\alpha}$ yields $F_{\rm out}(E)\propto E^{-\alpha+\Delta \alpha}$ outside
	the Milky Way as $F_{\rm out}=F_{\rm obs}(1+\Delta F)^{-1}$. Due to the
	strong increase (decrease) of $\Delta F$ at $(5-10)\,\text{EeV}$
	($(10-20)\,\text{EeV}$), especially for the field of view of Auger, the
	spectral index differs by $\Delta\alpha\simeq-0.2$ ($\Delta\alpha\simeq
	0.15$). Note that in the field of view of TA $\Delta\alpha$ is about a factor
	of two smaller.  In general, both experiments underestimate the flux outside
	the Milky Way, except around \SI{10}{\EeV} where they overestimate it.\footnote{At
	about $1\,\text{EeV}$, they might also observe an amplified flux, but this
depends strongly on the direction of the dipole at these energies which is
currently not known.} Above $10\,\text{EeV}$ this effect increases with energy
and the spectral behavior of $F_{\rm out}$ becomes harder than the observed
one. Comparing $\Delta F$ at energies $<10\,\text{EeV}$ for Auger's field of
view with the one for TA's field of view, the middle and right
Fig.~\ref{modObsData} expose that the flux suppression for Auger is up to
$(2-4)\,\%$ larger than for TA. This can be compared to the approx.~10\%
difference between the full-Sky spectra measured by Auger and TA below the
ankle~\cite{AbuZayyad2019}.  In the common observation band the difference
$(\Delta F_{\rm Auger}-\Delta F_{\rm TA})$ necessarily vanishes, but still
there is a significant UHECR flux bias, in particular at some tens of EeV.
However, the flux discrepancy between Auger and TA above $30\,\text{EeV}$ can
not be fully explained by the modification of the UHECR flux with the Galactic
magnetic field model JF12.  Note that the previously described effect of
$\Delta F$ increases by about a factor of two in the case of
$\lambda=100\,\text{pc}$. 

\begin{figure}[htbp]
\centering
\includegraphics[width=.4\linewidth]{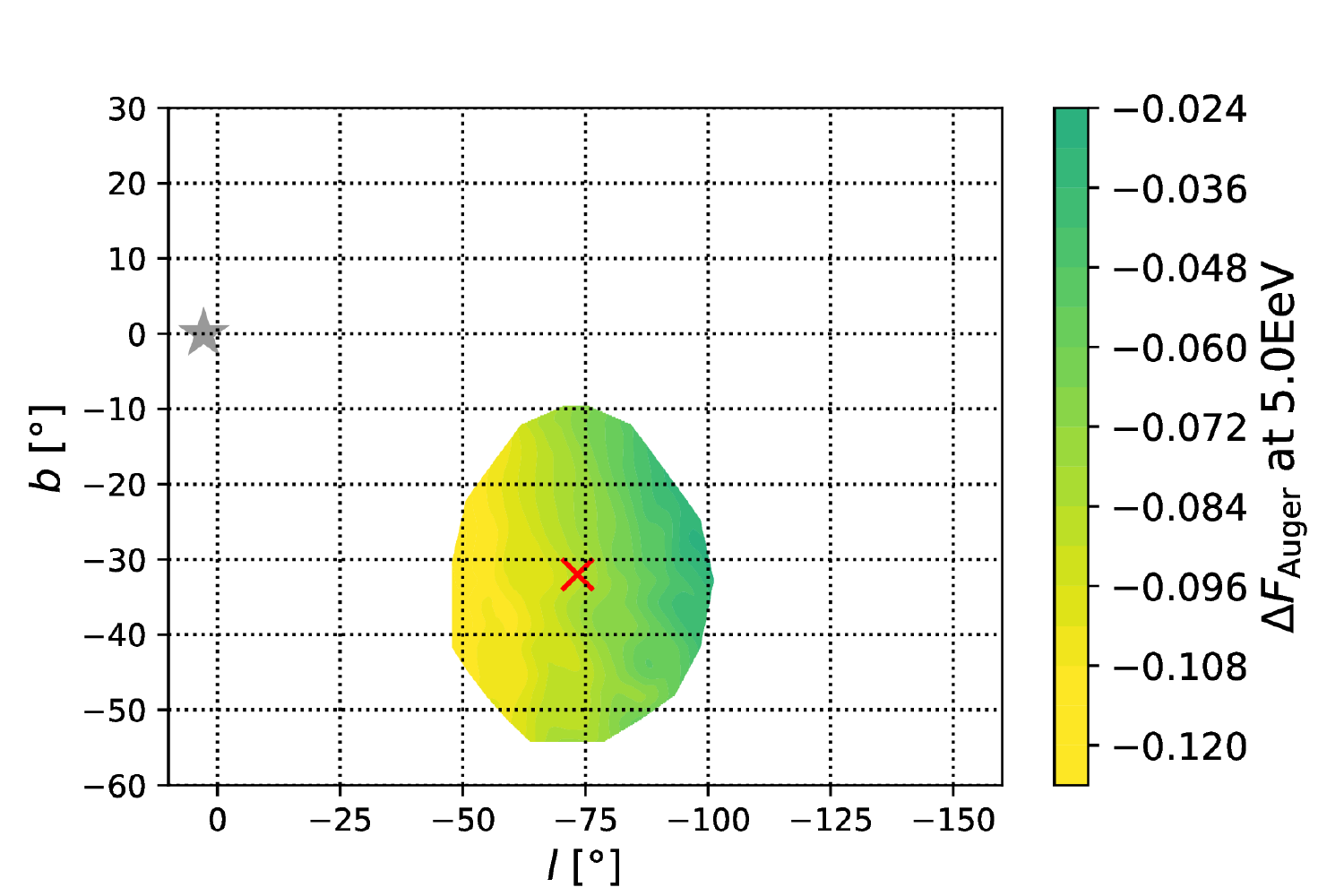}
\includegraphics[width=.4\linewidth]{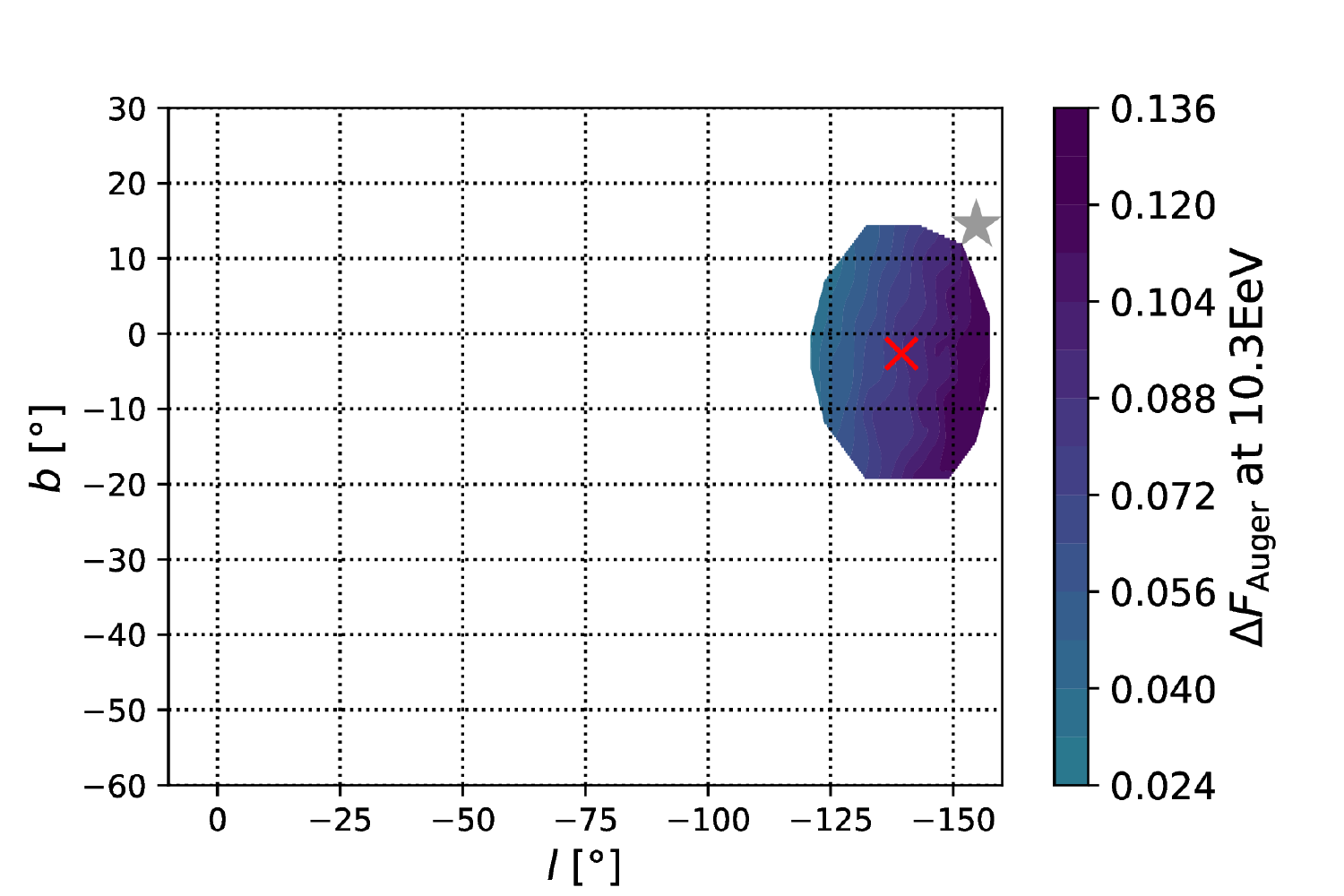}
\includegraphics[width=.4\linewidth]{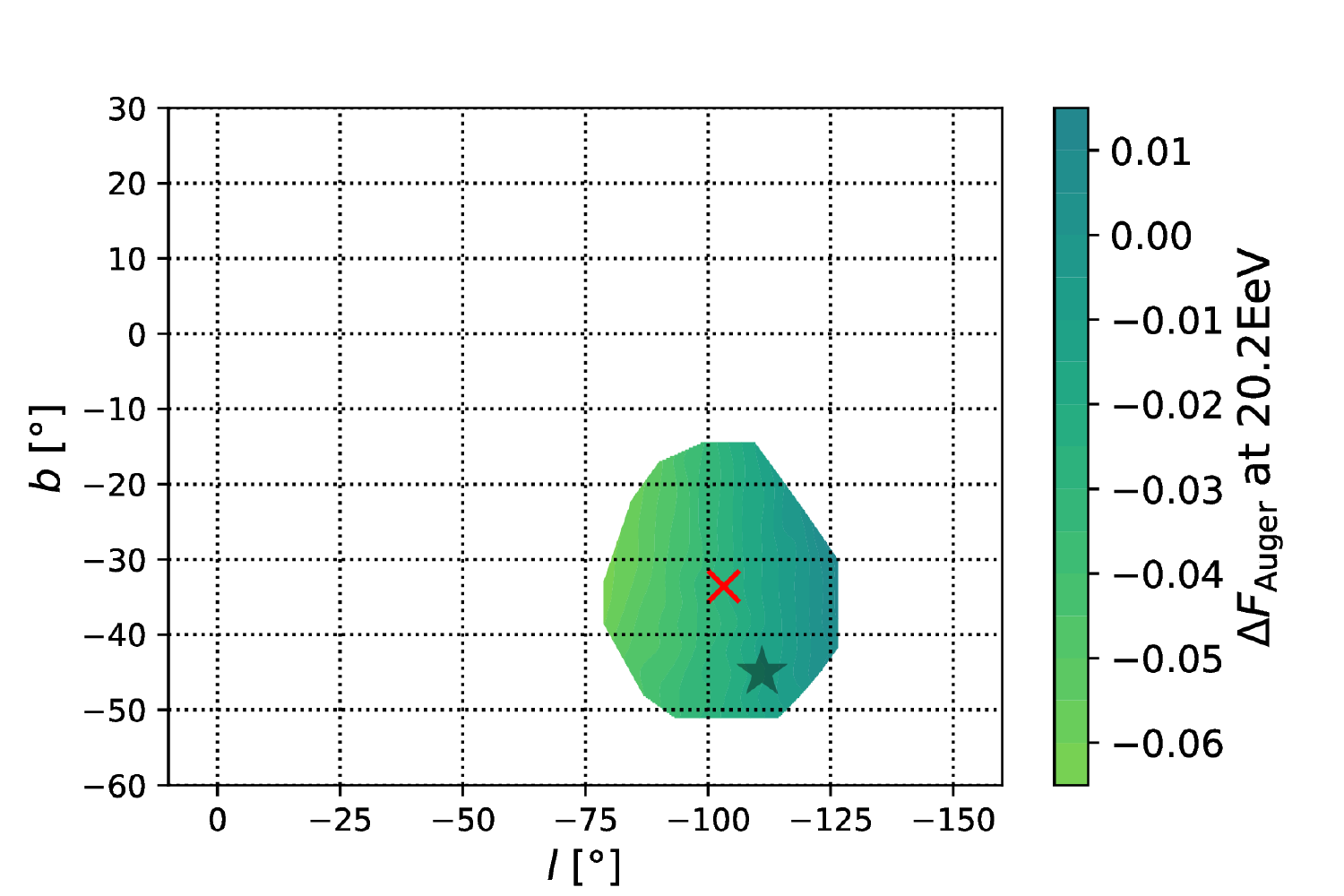}
\includegraphics[width=.4\linewidth]{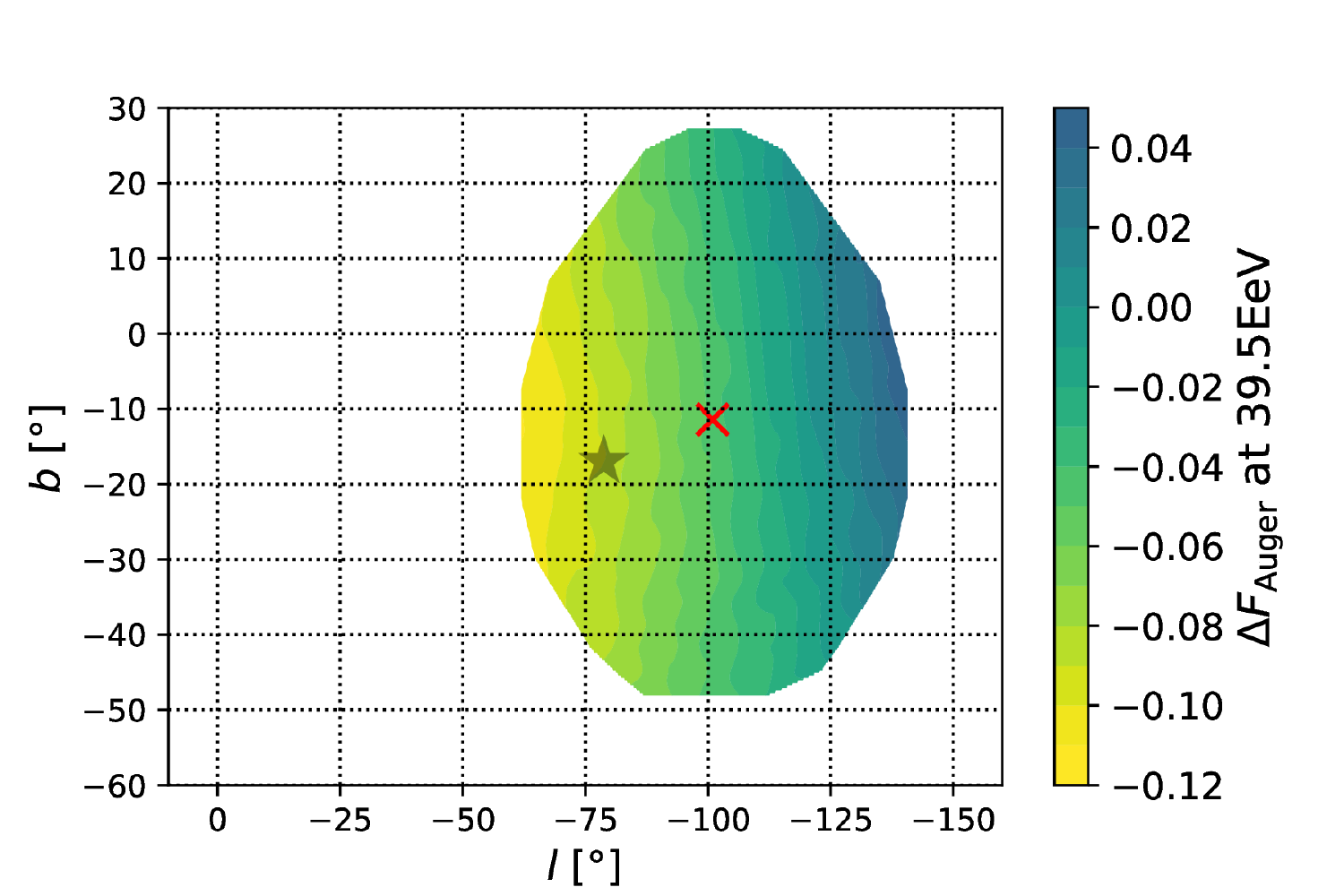}
\caption{The directional dependence of the UHECR flux bias for Auger's field of
view. The red cross indicates the observed mean dipole direction and the black
star refers to the corresponding dipole direction outside the Galaxy. Hereby,
the hadronic interaction model 'EPOS-LHC' as well as a coherence length of
$\lambda=60\,\text{pc}$ has been used.}
\label{obsErrIncl}
\end{figure}

The directional dependence of $\Delta F_{\rm Auger}$, as given in
	Fig.~\ref{obsErrIncl}, shows that a smaller longitude $l$ of the observed
	dipole --- within the range of the uncertainty --- leads to a stronger
	suppression by about a factor of two, and vice versa in the case of the
	amplification at $10.3\,\text{EeV}$. In addition, the shift of the mean
	dipole direction by the Galactic magnetic field suggests that there are
	different source directions: At $5\,\text{EeV}$ the dipole direction
	$(l,\,b)_{\rm out}$ outside the Galaxy is likely close to the Galactic
	center, whereas at higher energies a rather high Galactic longitude is
	favored. In particular, at energies $E\geq20\,\text{EeV}$ the dipole
	direction outside the Galaxy might stay the same if we account for the
	observational uncertainties. Note, that $(l,\,b)_{\rm out}$ significantly
	depends on the used hadronic interaction model, especially at low energies,
	as shown in Table \ref{DipParameters}.

\subsection{Constraints on the chemical composition}
\begin{figure}[htbp]
\centering
\includegraphics[width=.31\linewidth]{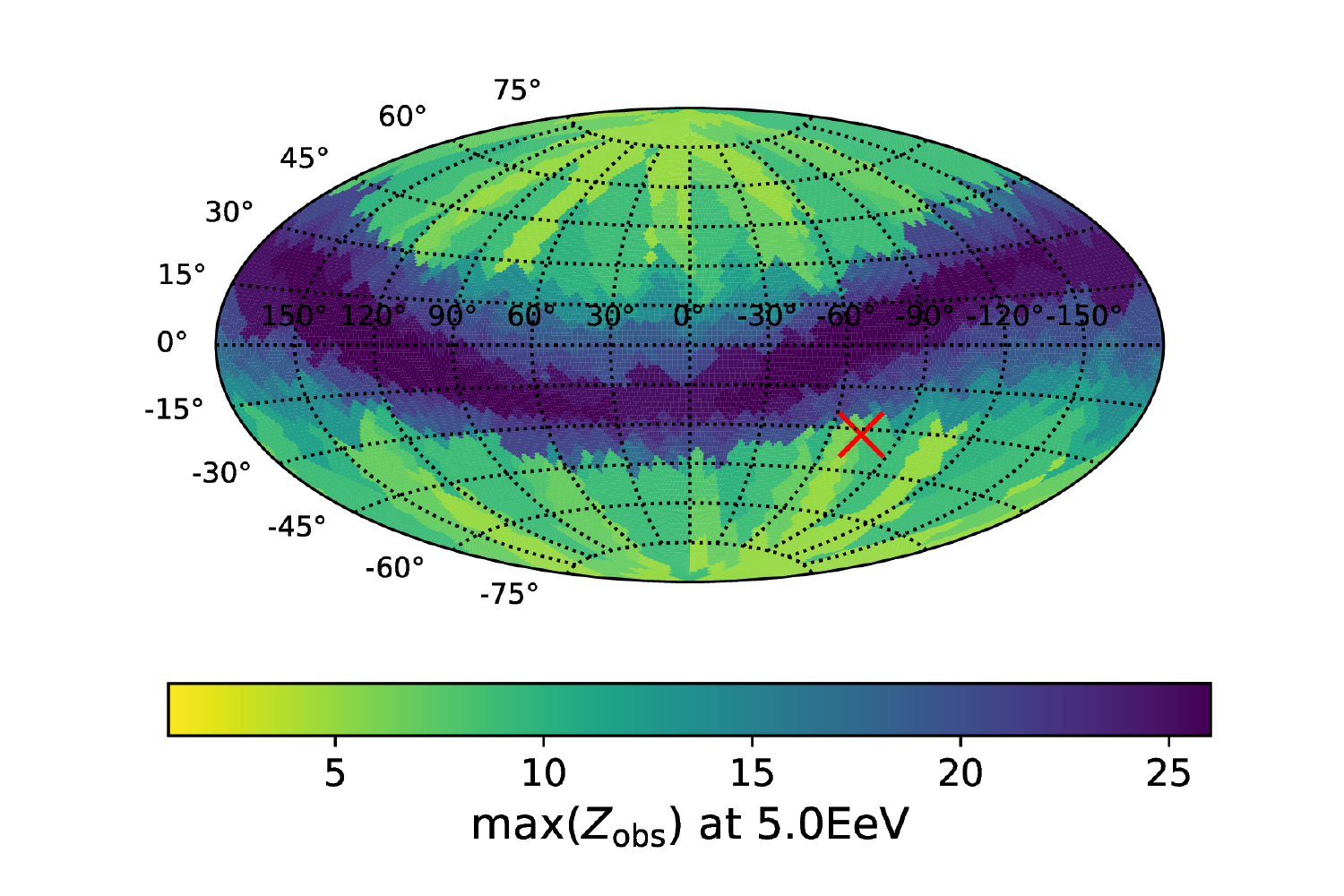}
\includegraphics[width=.31\linewidth]{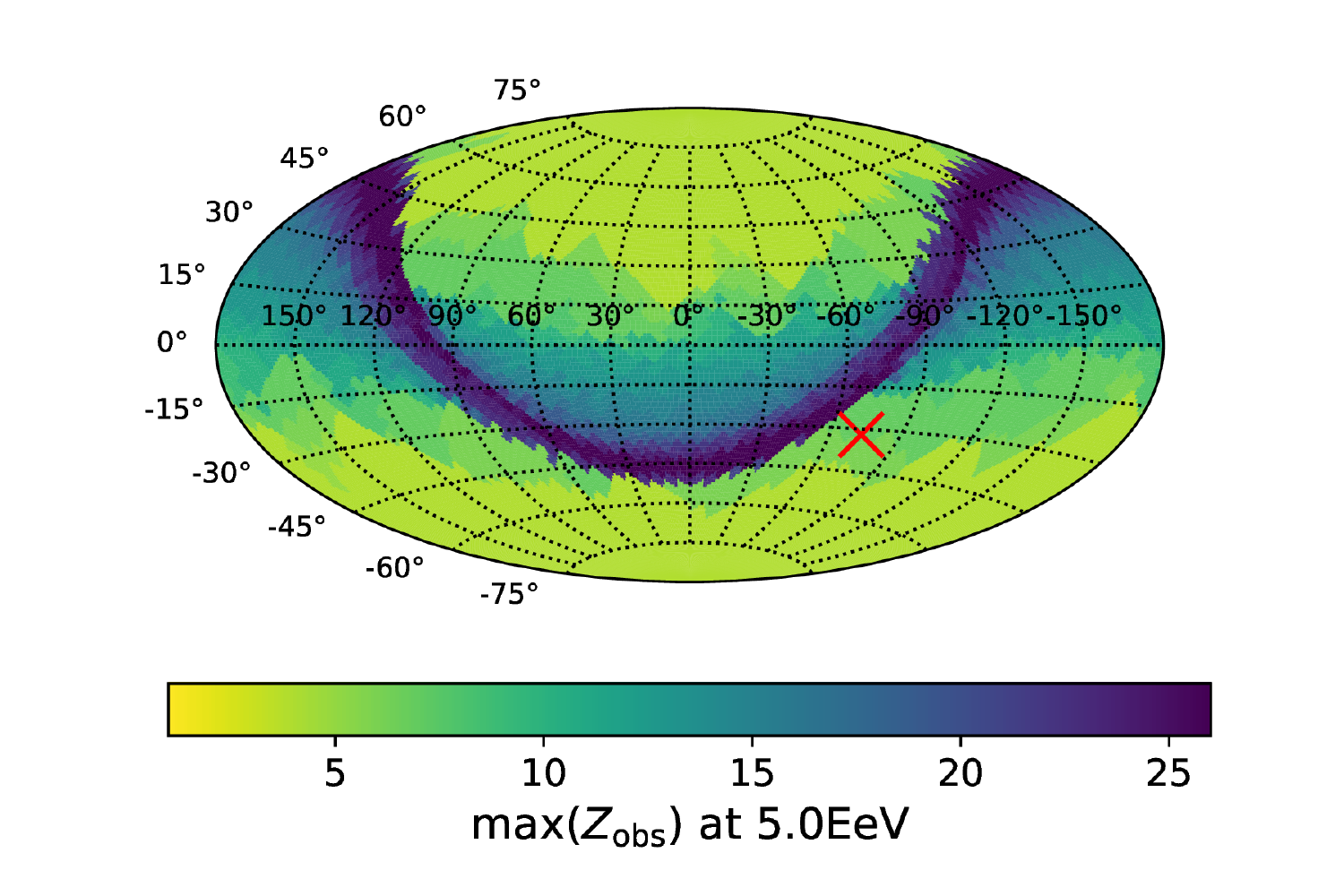}
\includegraphics[width=.31\linewidth]{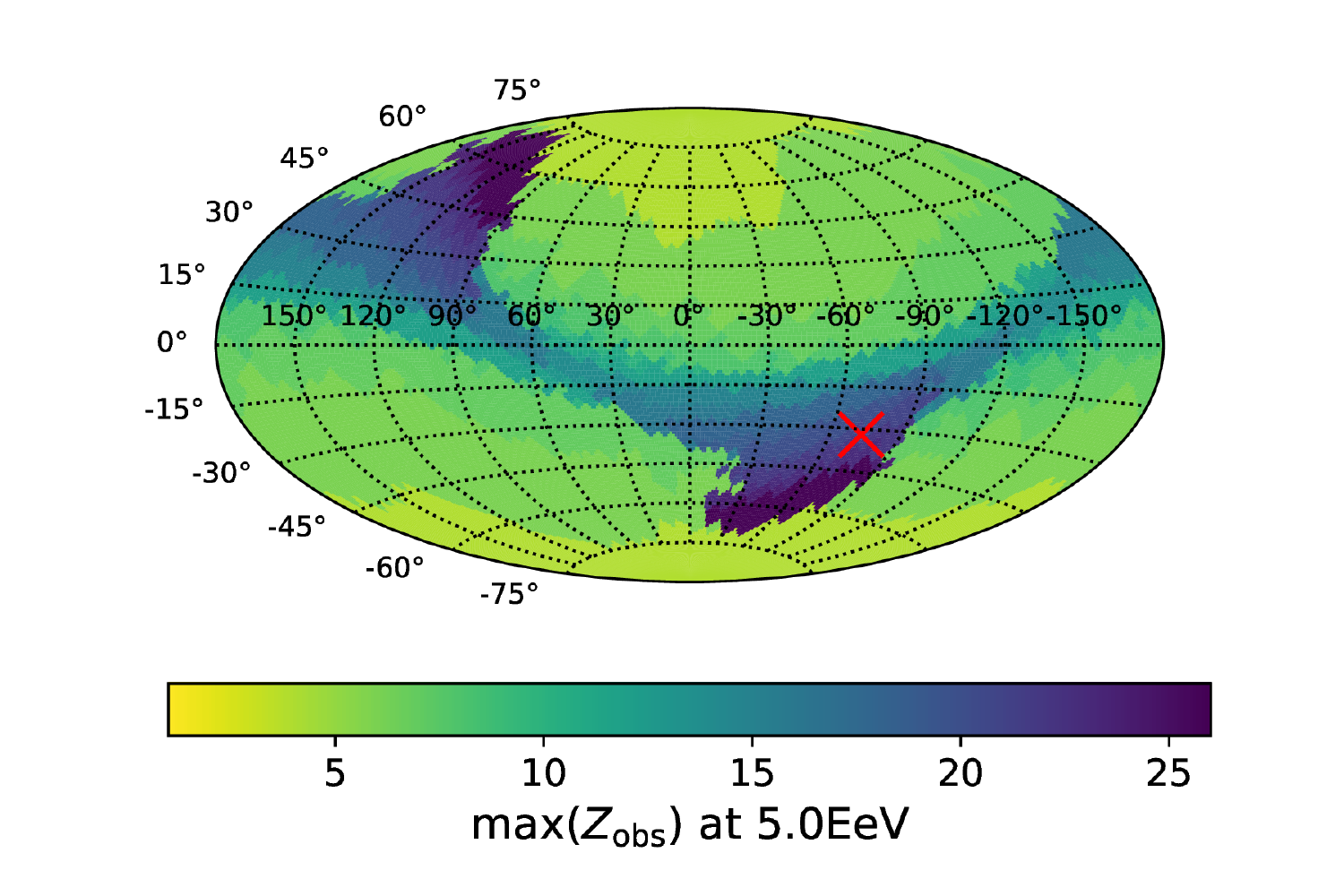}
\includegraphics[width=.31\linewidth]{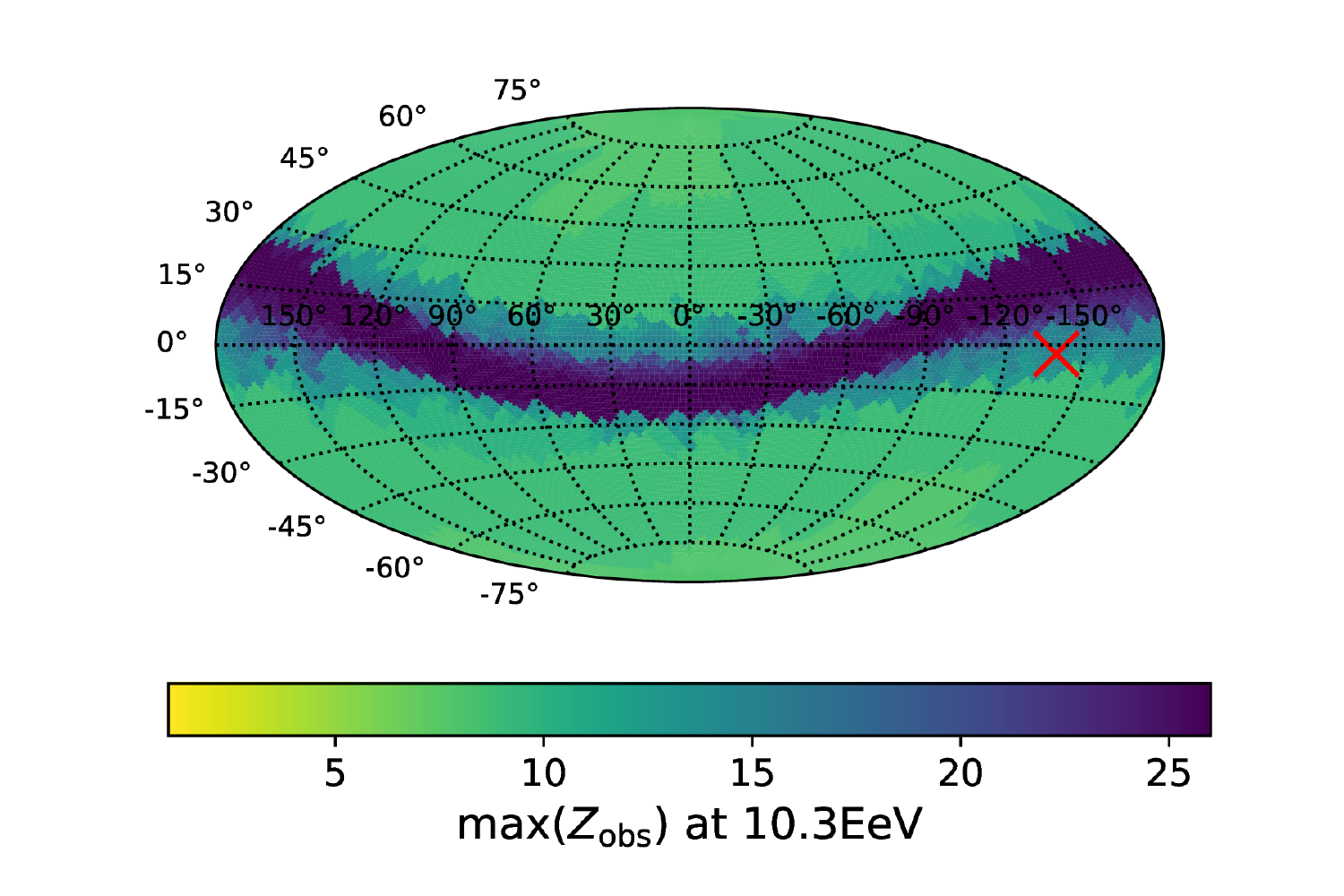}
\includegraphics[width=.31\linewidth]{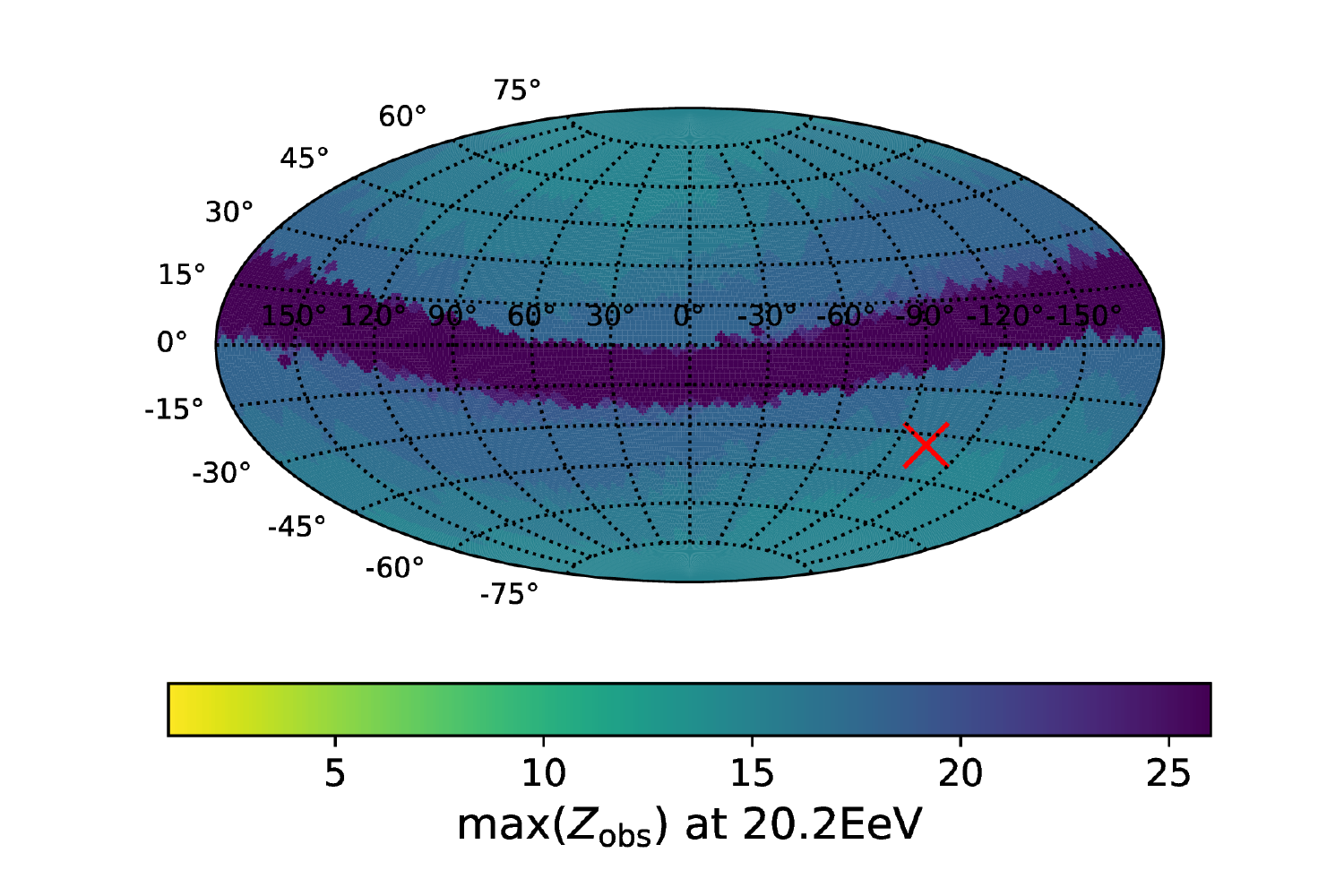}
\includegraphics[width=.31\linewidth]{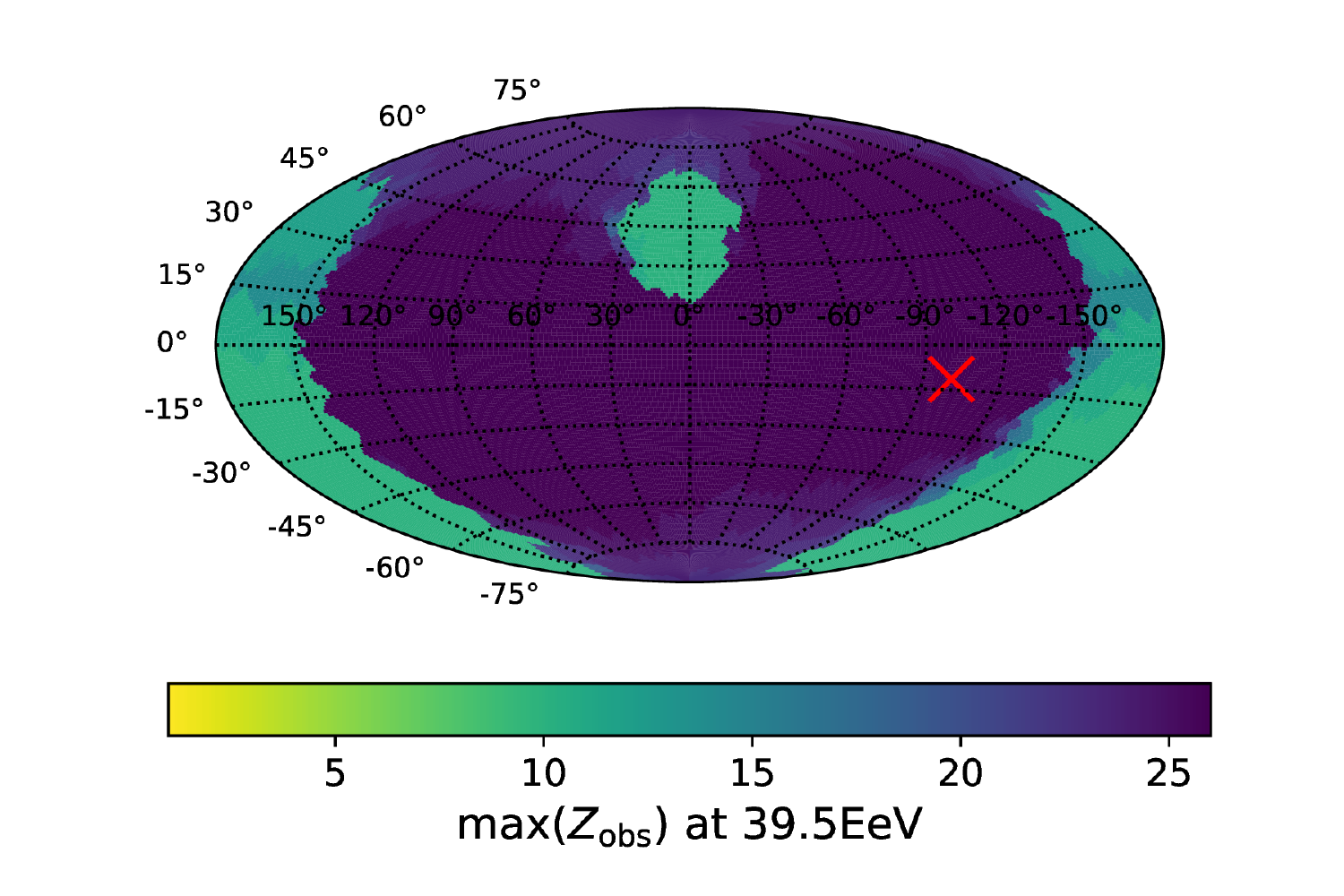}
\caption{The maximal charge number of UHECRs dependent on the direction of the
dipole. The red cross indicates the observed mean dipole direction by Auger.
\textbf{Upper panel:} The dipole at $5\,\text{EeV}$ using a coherence length of
the Galactic magnetic field of $\lambda=10\,\text{pc}$ (\textbf{left}),
$\lambda=60\,\text{pc}$ (\textbf{middle}) and $\lambda=100\,\text{pc}$
(\textbf{right}). \textbf{Lower panel:} The dipole at higher energies using
$\lambda=60\,\text{pc}$.}
\label{maxZ}
\end{figure}

Using the constrain on the maximal dipole amplitude, i.e.\ $d_{\rm out}\leq 1$,
we use the Auger data of the observed dipole amplitude and the corresponding
mean energy as given in Table \ref{DipParameters}, and determine the maximal mean
charge number $\text{max}(Z)$ dependent on the dipole direction. Hereby, we
compute the necessary rigidity at the observed cosmic ray energies and dipole amplitudes
in order to obtain $d_{\rm out}\geq 1$.
Fig.~\ref{maxZ} indicates that the Galactic magnetic field yields
an intriguing method to constrain the chemical composition of UHECRs for
certain directions of the dipole that only depends on the Galactic magnetic
field model and not on the hadronic interaction model.
At low energies the observed dipole --- in particular its mean
	direction --- points towards a direction that constrains the maximal charge
	number of the UHECRs. Hence, for a coherence length of some tens of pc, the
	mean chemical composition at $5\,\text{EeV}$ is at most constituted by CNO
	nuclei, whereas at $10\,\text{EeV}$ and $20\,\text{EeV}$ it could already be
	given by Si nuclei. Note, that in all cases the mean dipole direction is
	close to the narrow band that does not constrain the composition. Though,
	especially at $5\,\text{EeV}$ the coherence length of the Galactic magnetic
	field changes this band, and for $\lambda=100\,\text{pc}$ the constrain is
	significantly weaken.
At the highest energies this method does not draw any constraints on the composition of UHECRs. 

\section{Conclusions}
Using the Galactic lenses from the publicly available software package of
CRPropa3, we investigated the impact of dipole and quadrupole anisotropies on
the UHECR flux at Earth for the JF12 Model for the Galactic magnetic field. The
flux modification $\Delta F$ in case of a quadrupole anisotropy is smaller by
about a factor of two compared to the dipole anisotropy, however, the
quadrupole amplitude is more reduced by the Galactic magnetic field than the
dipole amplitude. Further, this effect allows to draw some compelling limits on
the maximal observed amplitude of the anisotropy, in particular at small
rigidities.

Considering the observed anisotropy level of a few percentages, it is shown
that in general, a modification of the UHECR flux by more than $10\%$ is
possible for rigidities $R < 5\,\text{EV}$ and certain direction of the
anisotropy, if we take the reduction of the anisotropy amplitude into account.
Further, this modification by the Galactic magnetic field can in principle also
produce differences in the measured UHECR spectrum of different experiments,
and thus explain some of the differences between the measurements of the Auger
and TA collaborations.
Finally, we account for the observed chemical composition, as well as the
observed  dipole direction and amplitude, showing that the UHECR spectrum can
be biased by more than $10\,\%$ for the hadronic interaction model 'EPOS-LHC'.
Though, for all interaction models, this effect yields at a few EeV a
suppression of the flux that is larger than the observational uncertainties. If
we account for the observational uncertainties of the dipole direction, which
allows directional shifts of several tens of degree, this effect can even be
increased by about a factor of two. In total, this work demonstrates that the
bias by the Galactic magnetic field impacts the interpretation of spectrum and
composition data.  So, it can lead to a suppression at the median energy of
$5\,\text{EeV}$ of up to $12\,\%$ and an amplification at $10\,\text{EeV}$ of
up to $13\,\%$, which implies a change of the UHECR flux spectrum by $\Delta
\alpha\simeq 0.2$ at most.

Further, we demonstrated that the amplitude and the mean energy of the observed
dipole can be used to draw constraints on the maximal mean charge number of the
UHECRs. Considering the mean direction again the charge number can be
constrained to $Z\lesssim 7$ at $5\,\text{EeV}$, if the turbulent component of
the Galactic magnetic field has a coherence length of about
$\lambda=60\,\text{pc}$. Hence, if the UHECRs at this energy are on average
composed of CNO nuclei there can only be a single dominant source outside the
Galaxy. Otherwise the dipole amplitude $d_{\rm out}\ll 1$, so that the
amplitude at Earth $d\ll d_{\rm obs}$. This consequence is in good agreement
with previous investigations on the reduction of the degree of anisotropy by
the JF12 model \cite{Farrar_2019}.

However, this effect does not fully resolve any of the discrepancies between
the observations of different experiments, since the flux difference $(\Delta
F_{\rm Auger}-\Delta F_{\rm TA})_{\rm d}$ necessarily vanishes in the common
observation band. Still, at several energies the bias $\Delta F$ can be at
about the same order as the reported difference in the case of a heavy
composition of UHECRs.

It is intriguing that with the observed dipole some difference between the
observed spectra of the experiments is expected.  For updated dipole directions
based on additional data, or the discovery of a significant quadrupole
anisotropy, the Galactic magnetic field bias can become a major issue.  Note,
that these outcomes strongly depend on the used Galactic magnetic field model,
and due to the current issues with this model, we strongly encourage to repeat
this analysis with improved magnetic field models.

\acknowledgments

Some of the results in this paper have been derived using the software packages
Numpy \cite{vanDerWalt2011}, Matplotlib \cite{Hunter:2007} and HEALPix/ healpy
\cite{2005ApJ...622..759G}. TW acknowledges supported by DFG grant WI~4946/1-1.

\bibliographystyle{JHEP}
\addcontentsline{toc}{section}{Bibliography}
\bibliography{references}

\end{document}